\documentclass{aastex7}

\usepackage{subcaption}
\usepackage{graphicx}

\begin{document}

\title{3D MHD wave propagation and energy transport in a simulated solar vortex}

\author[orcid=0000-0002-3814-4232,gname=Samuel J.,sname=Skirvin]{Samuel J. Skirvin}
\affiliation{Plasma Dynamics Group, School of Electrical and Electronic Engineering, University of Sheffield, Sheffield, S1 3JD, UK}
\email[show]{s.skirvin@sheffield.ac.uk}  

\author[orcid=0000-0002-0893-7346,gname=Viktor, sname=Fedun]{Viktor Fedun}
\affiliation{Plasma Dynamics Group, School of Electrical and Electronic Engineering, University of Sheffield, Sheffield, S1 3JD, UK}
\email{v.fedun@sheffield.ac.uk}  

\author[orcid=0000-0002-9546-2368,gname=Gary, sname=Verth]{Gary Verth}
\affiliation{Plasma Dynamics Group, School of Mathematical and Physical Sciences, University of Sheffield, Sheffield S3 7RH, UK}
\email{g.verth@sheffield.ac.uk}  

\author[orcid=0000-0002-3066-7653,gname=Istvan, sname=Ballai]{Istvan Ballai}
\affiliation{Plasma Dynamics Group, School of Mathematical and Physical Sciences, University of Sheffield, Sheffield S3 7RH, UK}
\email{i.ballai@sheffield.ac.uk}

\begin{abstract}

Magnetic flux tubes in the presence of background rotational flows are abundant throughout the solar atmosphere and may act as conduits for MHD waves to transport energy throughout the solar atmosphere. Here we investigate the contribution from MHD waves to the Poynting flux in a 3D numerical simulation of a realistic solar atmosphere, modelling a structure resembling a solar vortex tube, using the PLUTO code in the presence of different plasma flow configurations. These simulations feature a closed magnetic loop system where a rotational flow is imposed at one foot-point in addition to photospheric perturbations acting as a wave driver mimicking those of p-modes. We find that a variety of MHD waves exist within the vortex tube, including sausage, kink and torsional Alfv\'{e}n waves, owing to the photospheric wave driver and the nature of the rotational flow itself. We demonstrate how the visual interpretation of different MHD modes becomes non-trivial when a background rotational flow is present compared to a static flux tube. By conducting a simulation both with and without the rotational plasma flow, we demonstrate how the perturbed Poynting flux increases in the presence of the rotational flow as the waves transport increased magnetic energy. We attribute this increase to the dynamical pressure from the rotational flow increasing the plasma density at the tube boundary, which acts to trap the wave energy more effectively inside the vortex. Moreover, we demonstrate how the Poynting flux is always directed upwards in weakly twisted magnetic flux tubes.

\end{abstract}



\section{Introduction} \label{sec:intro}

The solar magnetic field, coupled with dynamical plasma motions at the photosphere are known to play a major role in the transport of energy to the solar corona which is ultimately responsible for heating coronal plasma to millions of degrees \citep{Klimchuk2006}. One of the possible ways how energy can be transported is through the propagation of magnetohydrodynamic (MHD) waves in various magnetic structures, supported by gradients in the magnetic field or other plasma diagnostics \citep{vanDoorsselaere2020SSRv}. 

Solar vortices are ubiquitous throughout the solar atmosphere \citep{jess2015, Tziotziou2023SSRv, Skirvin2023_JET} and small-scale swirling motions occur naturally in the photosphere within the intergranular lanes \citep{Bonet2008, Shelyag2011vorticity, Giagkiozis2018} in addition to magnetic structures which display rotational behaviour \citep[e.g.][]{Curdt2011, Su2014, Sharma2018, Skirvin2023_JET}. Photospheric swirling motions have been suggested as a source of incompressible torsional Alfv\'{e}n waves in the solar atmosphere \citep{fed2011, Morton2013, Yadav2021, Jess2023}. Moreover, solar vortex tubes have been reported to harbour oscillations which may be interpreted as MHD waves \citep{Tzi2018, Tzi2020,  Murabito2020}. Whilst it is becoming more evident that rotating structures can support MHD waves in the solar atmosphere, there are few studies investigating the potential for solar vortex tubes to transport MHD wave energy or how MHD waves may manifest themselves in such structures \citep{Skirvin2023rotflow}.

Recent analytical work \citep{Skirvin2024Poynting} has demonstrated that MHD waves carry an increased amount of field aligned Poynting flux in the presence of background rotational flows and that the flux increase is dependent upon the amplitude ratio between the strengths of the flow and the perturbation. Additionally, numerical simulations have shown that the Poynting flux is enhanced in regions where solar vortices are present, either in the form of solar tornadoes or small-scale vortices within the solar plage \citep{Yadav2020, Yadav2021, Finley2022, Kuniyoshi2023}. Numerical simulations of a magnetic tornado structure in the solar atmosphere \citep{Wedemeyer2012} concluded that vortices contain 14 kW m$^{-2}$ of vertical Poynting flux with a net upward flux of 400~W m$^{-2}$. However, in these studies, the authors did not elaborate on the possibility for the enhanced Poynting flux to be transported from supported MHD wave propagation within the vortex structure, or attempt to separate the vertical Poynting flux associated with the vortex structure itself from the Poynting flux carried by MHD waves. In both numerical magnetoconvection simulations and solar observations, it is difficult to separate the background plasma motions from any perturbations (e.g. waves) from the full field vectors. Our work presents a first attempt at conducting full 3D modelling of MHD waves in a magnetic flux tube in the presence of a background vortical flow, with the aim to dissect the properties and energy associated with MHD wave propagation.

Our paper is structured as follows: in Section \ref{sec:methods} we describe the numerical model used in the current study and the creation of a 3D solar vortex tube, in addition to introducing the modelled wave driver. Section \ref{sec:results} describes the results of the current study investigating the MHD waves present in the vortex tube and their different generation mechanisms. Furthermore we present results on the Poynting flux in the vortex tube and the contribution to this flux from MHD waves. Section \ref{sec:conclusions} summarises the results and outlines avenues for future work.

\section{Methods} \label{sec:methods}
\subsection{Model} \label{subsec:model}

\begin{figure*}[ht!]
\centering
\begin{subfigure}{0.32\textwidth}
    \includegraphics[width=\textwidth]{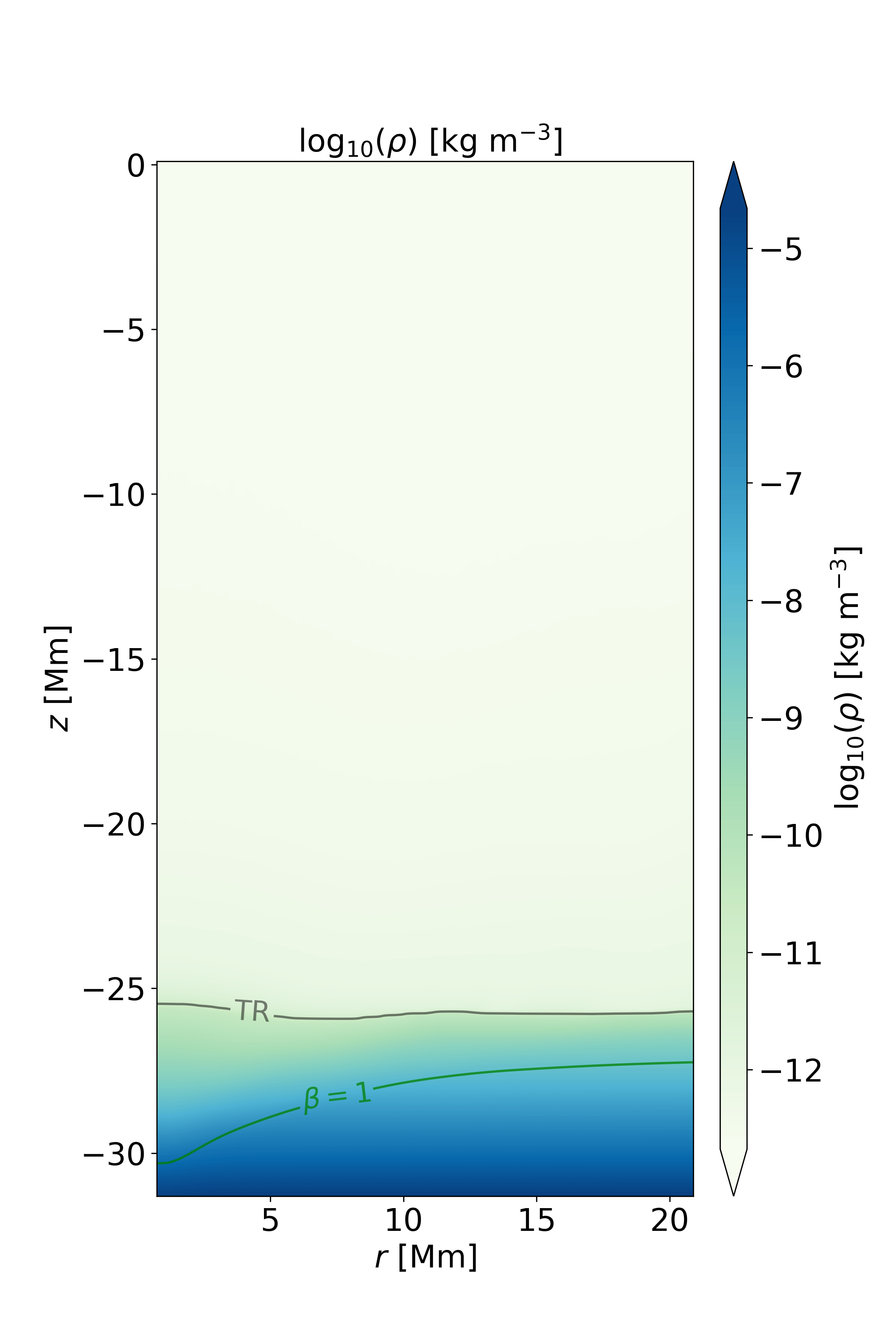}
    \caption{}
    \label{fig:equilibrium_density}
\end{subfigure}
\begin{subfigure}{0.32\textwidth}
    \includegraphics[width=\textwidth]{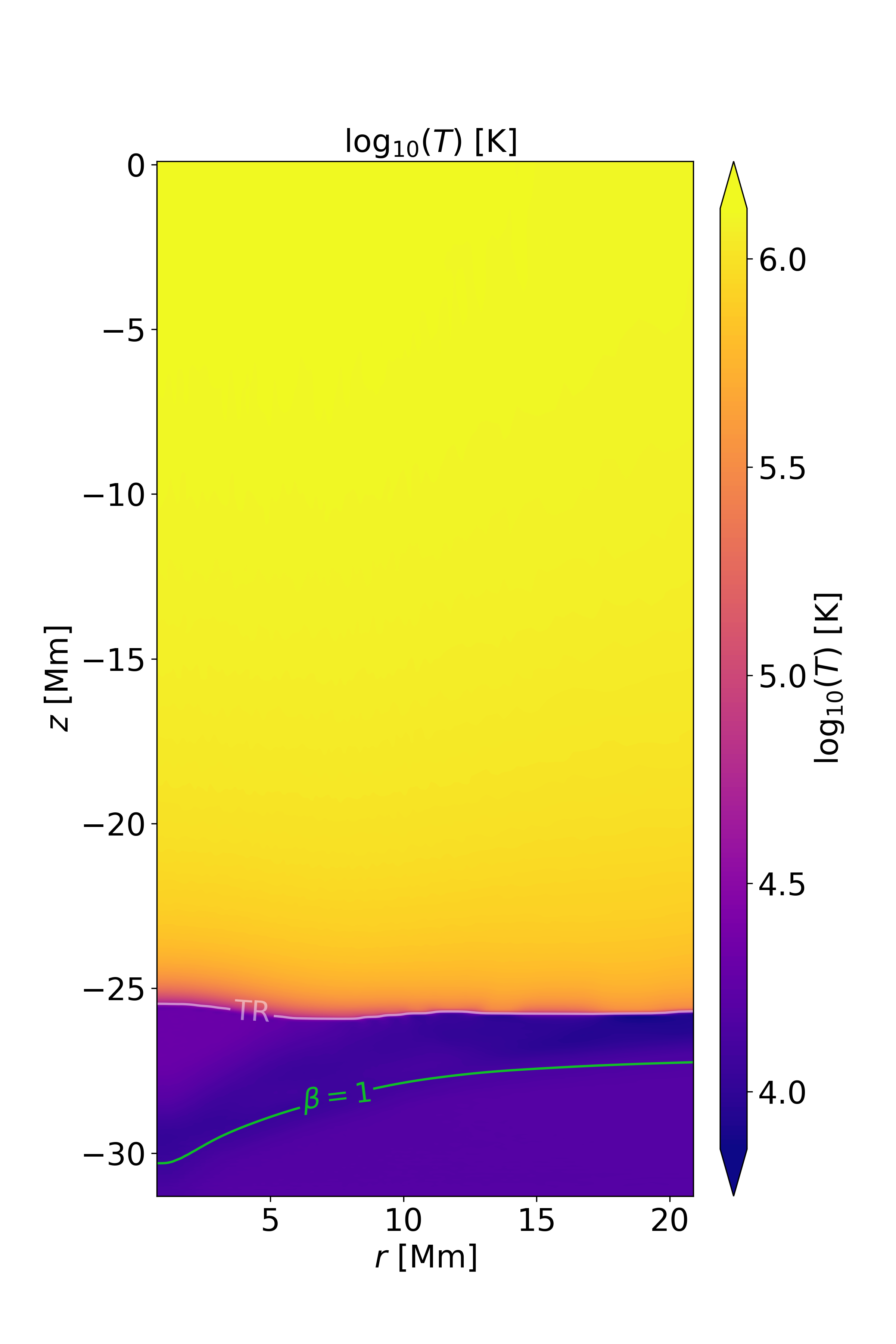}
    \caption{}
    \label{fig:equilibrium_temp}
\end{subfigure}
\begin{subfigure}{0.32\textwidth}
    \includegraphics[width=\textwidth]{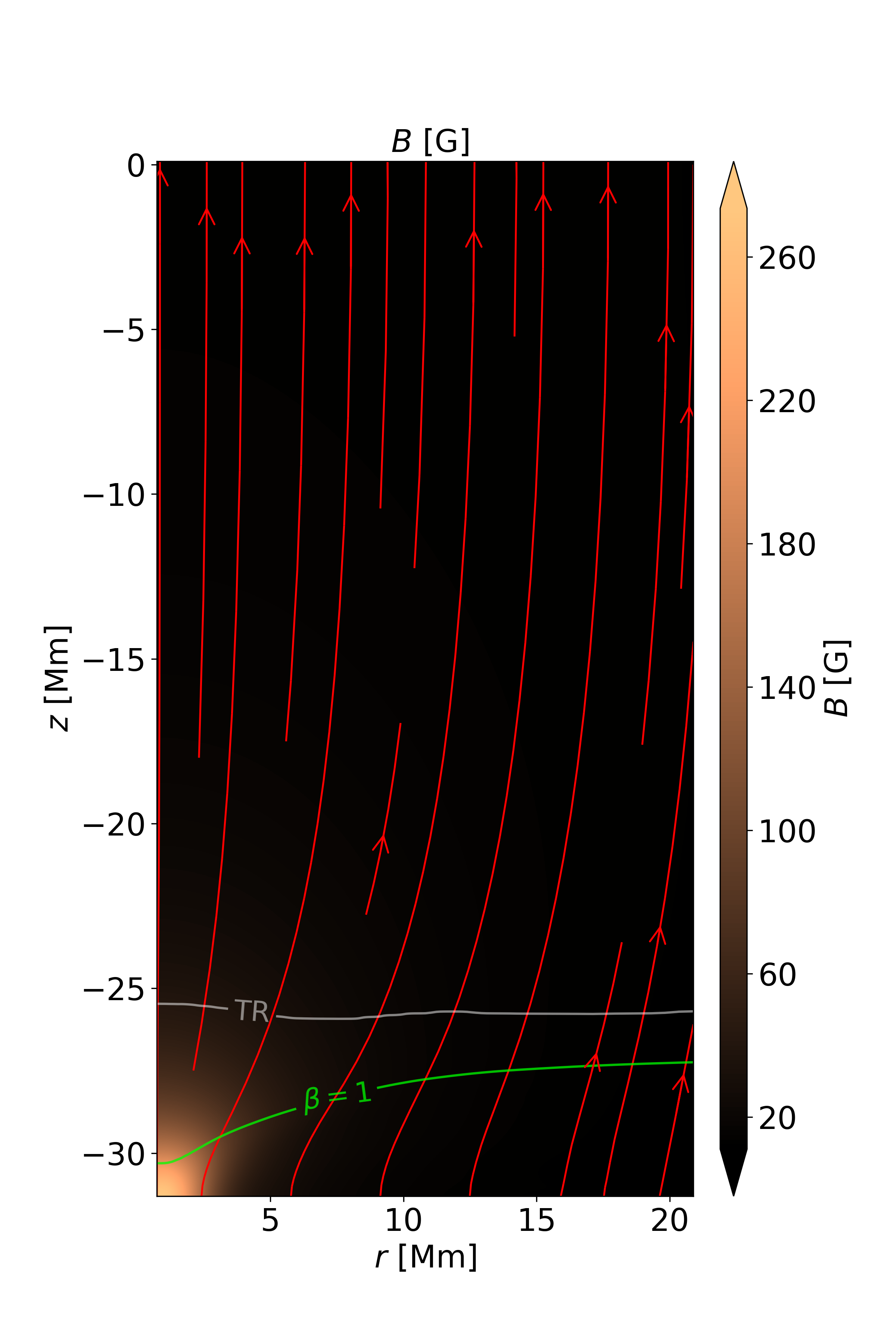}
    \caption{}
    \label{fig:equilibrium_magfield}
\end{subfigure}
\caption{The equilibrium configuration displaying (a) the plasma density, (b) plasma temperature and (c) magnetic field strength and structure in the 3D loop model. This corresponds to a snapshot at the initial stage of the simulation before any flow or wave driver is introduced and is taken at an azimuthal slice of $\varphi=\pi/2$. The transition region (TR) and plasma-$\beta=1$ contours are also highlighted in each plot. This figure shows a subsection of the full numerical domain for illustration purposes.}
\label{fig:atmosphere}
\end{figure*}

The 3D numerical model adopted in this work was introduced by \citet{Reale2016} featuring a straightened, evacuated loop spanning from photosphere to photosphere in a cylindrical coordinate system $(r, \varphi,z)$. The simulation domain ranges from $0.73$ Mm to $41.01$ Mm in the radial direction, from $0$ to $\pi$ in the $\varphi$-direction, and from $-31.31$ Mm to $31.31$ Mm in the $z$-direction, with $192$ × $256$ × $768$ data points, respectively. The original thermodynamic model is adapted from \citet{Serio1981} for closed coronal loop models and the initial equilibrium is obtained by numerically relaxing a hydro-static model over time, with vertical and straight magnetic field, as outlined in \citet{Guar2014}. As a result of increased plasma and magnetic pressure at the centre of the loop, the magnetic field ultimately expands in the corona during the equilibration process. At the foot-points, the magnetic field has a maximum strength of $273$~G which decreases to $13$~G at the loop apex at $z = 0$~Mm (see Figure \ref{fig:equilibrium_magfield}), which is appropriate for quiet Sun conditions of the network field. Furthermore, the total magnetic field decreases with radial distance across the loop and its magnitude has a profile which is approximately Gaussian-shaped. For more information regarding the numerical domain and setup, we refer readers to earlier studies by \citet{Riedl2021, Skirvin2023ApJ, Skirvin2024PDs}. 

Gravity is incorporated into the model in the form:
\begin{equation}
    g(z)=-g_\odot \cos{\left(\frac{(z-z_0)\pi}{L}\right)},
\end{equation}
where $g_\odot=274$ m s$^{-2}$ is the gravitational acceleration at the solar surface, $z_0$ is the $z$-coordinate of the photosphere located at the base of our numerical domain, and $L = 61.61$~Mm is the total length of the loop. Taking a gravitational profile in this form results in a gravitational acceleration acting in the negative $z$-direction for $z<0$ and to the positive $z$-direction for $z>0$ with $g(0)=0$ m s$^{-2}$ at the loop apex. Gravity across the loop, as well as the loop curvature, are neglected.

The initial model is presented in Figure \ref{fig:atmosphere}, highlighting the plasma density and temperature, in addition to the plasma-$\beta = 1$ contour along with the position of the transition region (TR) which, for simplicity, is defined as the location where the plasma temperature $T = 40,000$~K. However as the TR has a finite width, the TR contour should be regarded as an approximation. The width of the TR is roughly $1$~Mm, which is broader than theoretically expected, but much narrower than some numerical models \citep[e.g.][]{Pelouze2023, karampelas2024}. Therefore, the TR will still act as a wave barrier, but may allow for increased transmission of energy flux than expected. We also display the strength and structure of the magnetic field in Figure \ref{fig:equilibrium_magfield}, which exhibits an expanding flux tube structure below the transition region where the density stratification is strong. The magnetic field is directed from the foot-point at the bottom boundary to the opposite foot-point at the upper boundary.

\subsection{Numerical Setup and Boundary Conditions}\label{subsec:setup}

As we are interested in studying MHD waves in the presence of rotational flows, a full 3D cylinder ($2\pi$ in the azimuthal direction) is modelled in the current work. This differs from previous simulations utilizing the same model, where only one quarter \citep{Reale2016,Riedl2021} or one half of the loop \citep{Skirvin2023ApJ, Skirvin2024PDs} were considered. Our simulation domain ranges from $0.73$ Mm to $41.01$ Mm in the radial direction, from $0$ to $2\pi$ in the $\varphi$-direction, and from $-31.31$ Mm to $31.31$ Mm in the $z$-direction, with $192$ × $512$ × $768$ data points, respectively. The loop axis is located at $r=0$~Mm, however, we do not simulate the domain close to this region due to the singularity at the origin of the domain. The loop apex is located at $z=0$~Mm. The numerical mesh is stretched in the radial and vertical directions, in certain regimes of the domain to ensure that increased resolution is obtained closer to the enhanced region of magnetic field, whereas decreased resolution is utilized in regions of low stratification such as the upper corona and at large radial distances away from the centre of the magnetic loop. The mesh and resulting resolution are uniform in the azimuthal direction. An exact description of the numerical grid and resolution in different regimes can be found in \citet{Riedl2021}.

The simulations are performed using the PLUTO code \citep{Mign2007, Mign2012, Mign2018}, where the ideal MHD equations are solved in 3D cylindrical coordinates using the Harten-Lax-Van Leer (HLL) approximate Riemann solver, with a piece-wise total variation diminishing (TVD) linear reconstruction method for the spatial integration. Field-aligned thermal conduction is included in the model, however, non-ideal effects such as viscosity, resistivity and radiative cooling are ignored. We utilize reflective boundary conditions for both boundaries in the $r$-direction and symmetric boundary conditions in the $\varphi$-direction. Additionally, we incorporate anti-symmetric boundaries for the upper boundary. At the lower boundary, the same boundary conditions are set; however the velocity, pressure and density are perturbed according to an analytical solution for a gravity-acoustic wave, given in Section \ref{subsec:driver}.

\subsection{Forming a 3D solar vortex tube}\label{subsec:vortex_formation}

To create a structure resembling a solar vortex tube, we introduce a velocity perturbation at the bottom boundary in the azimuthal component, $v_{\varphi}$, representing one foot-point of the loop. From this point forward we will refer to the magnetic structure as a vortex tube, which is created by modelling a rotational flow inside the magnetic loop. This perturbation introduces a twisting of the velocity at the photopsheric level. Due to the high plasma-$\beta$ here ($\beta \gg 1$) the magnetic field is tied to the plasma motions, as such, $B_{\varphi}$ are naturally generated at the lower boundary however are less pronounced at greater heights. These rotational perturbations are commonplace in the solar photosphere and naturally appear in the intergranular lanes \citep{Giagkiozis2018}. The torsional flow, in the horizontal plane at the bottom boundary, takes the form \citep{Magyar2021}:
\begin{equation}\label{vphi_rotflow}
    v_{\varphi} (r,t) = f(t) \cdot A \ e^{-((r-r_s)/r_b)^2},
\end{equation}
where $A$ represents the flow amplitude taken to be $1$~km s$^{-1}$, $r_s$ denotes the centre of the Gaussian and $r_b$ is the width of the flow in the horizontal plane. The width of the Gaussian in Equation (\ref{vphi_rotflow}) results in an azimuthal flow in a narrow ring of $r_b$ around $r_s$. We set $r_b=750$~km and $r_s=1.75$~Mm, which results in a vortex flow driven off-centre from the magnetic field enhancement at one foot-point. The time behaviour of the torsional flow is given by the function $f(t)$ and models a decaying vortex. The analytical form of this function is given by $f(t) = exp^{-(t/\tau)^8}$, where $\tau=270$~s represents the characteristic driving time \citep{Magyar2021}. 

\begin{figure}
\centering
\begin{subfigure}{0.45\textwidth}
    \includegraphics[width=\textwidth]{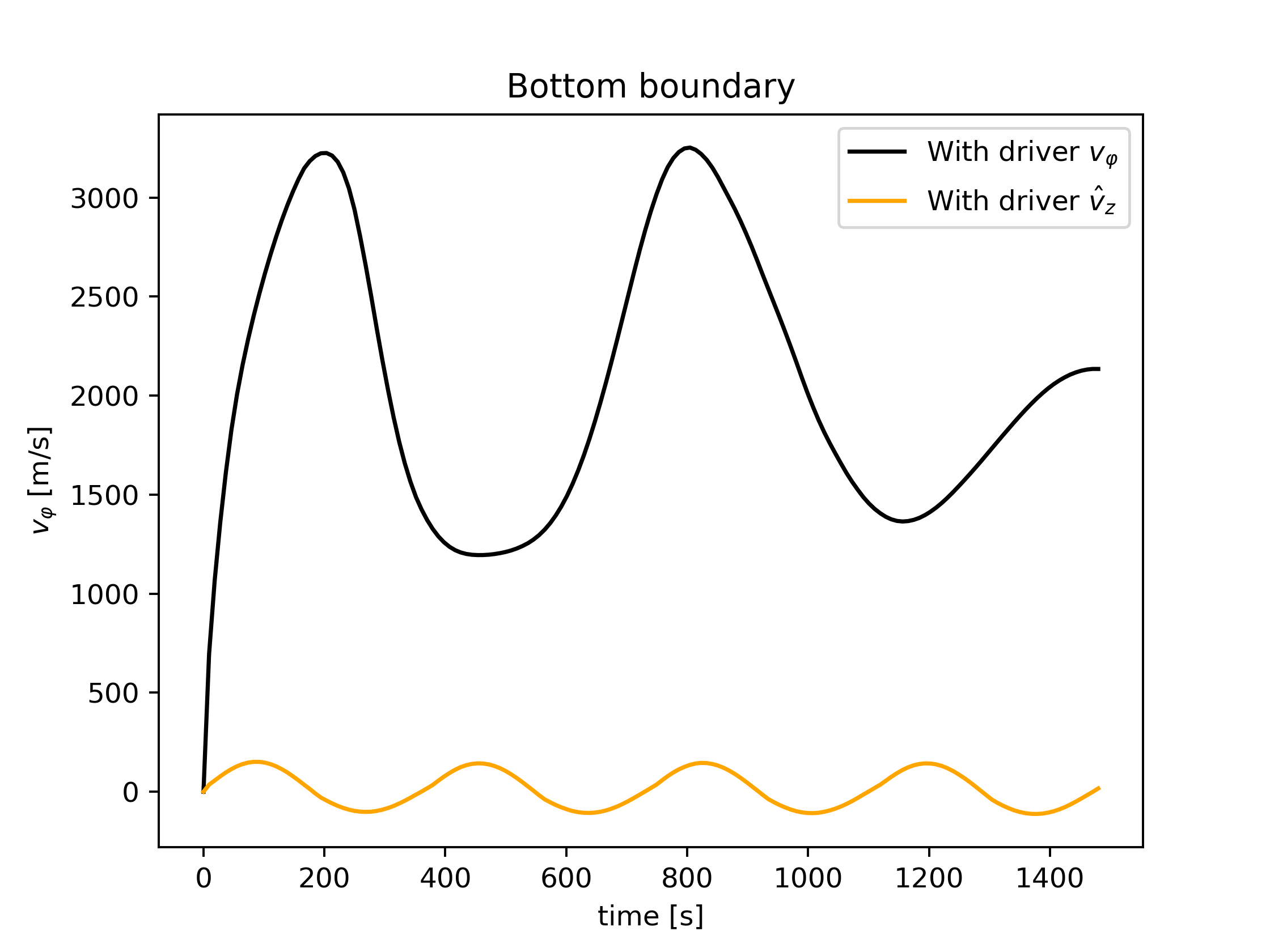}
    \caption{}
    \label{fig:vphi_with_time}
\end{subfigure}
\begin{subfigure}{0.45\textwidth}
    \includegraphics[width=\textwidth]{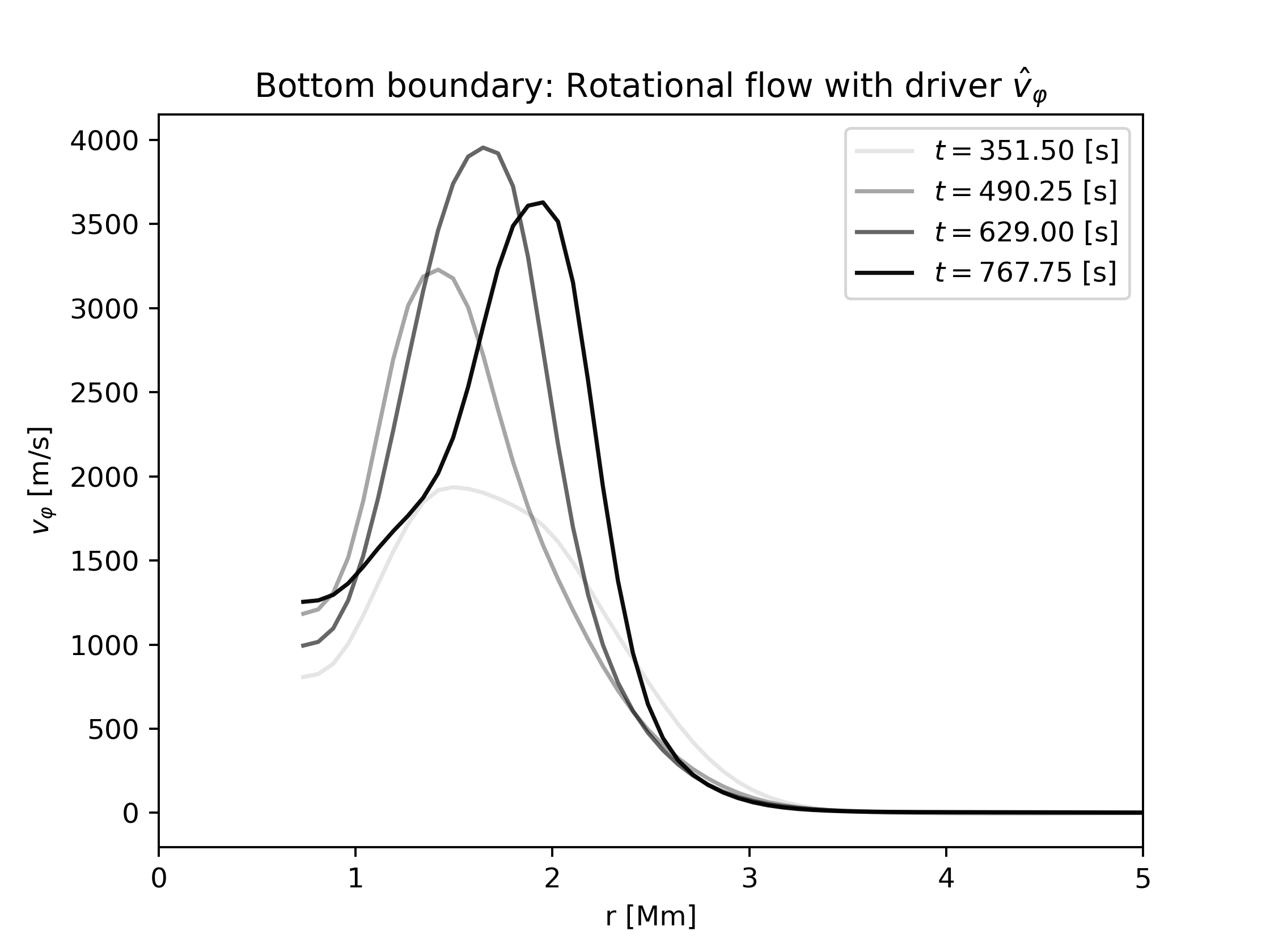}
    \caption{}
    \label{fig:vphi_with_radius}
\end{subfigure}
    
\caption{Plots at the bottom boundary of the domain showing (a) the temporal evolution of the total azimuthal velocity ($v_{\varphi}$) and perturbation of the vertical component of the velocity ($\hat{v}_z$) from the wave driver at $r=2.10$~Mm. Panel (b) highlights the spatial profile of the rotational flow applied at the bottom boundary at different time snapshots, depicted in the legend.}
\label{fig:bottomboundary}
\end{figure}

The temporal evolution of the velocity perturbation at the bottom boundary is displayed in Figure \ref{fig:vphi_with_time}. The profile of $v_{\varphi}$ demonstrates a unique periodicity which is not defined in Equation (\ref{vphi_rotflow}) and is a result of the boundary condition at the inner radial boundary. As the simulation domain does not cover the axis of the tube, the inner radial boundary is located at $r=0.71$~Mm with a reflective boundary condition. As a result, the azimuthal motions are tied to this location and experience a reflection as the vortex begins to decay. Whilst this physical effect is not intended in the simulation, it can perform to mimic the temporal behaviour of a realistic solar vortex within intergranular lanes in the photosphere, which constantly experiences buffeting motions from nearby granules affecting its rotational speed. This effect is only present in the radial dependence of the flow as Equation (\ref{vphi_rotflow}) is independent on $\varphi$ and the azimuthal boundary conditions are symmetric. Ultimately, this effect acts as an additional incompressible wave driver and is discussed in more detail in Section \ref{subsec:aziwavenumbers}. In fact, it is likely that the buffeting experienced by vortices in the intergranular lanes are also non-axisymemtric in nature (unlike the axisymmetric rotational flow considered in this work) which will act to excite higher order modes of different polarisation. Figure \ref{fig:vphi_with_time} also displays the amplitude of $\hat{v}_z$ associated with the photospheric wave driver as a function of time. The clear periodicity of the wave driver is observed, which is discussed in more detail in Section \ref{subsec:driver}.

Furthermore, the spatial distribution of the rotational flow applied at the bottom boundary is displayed in Figure \ref{fig:vphi_with_radius} for four time snapshots. The location of the inner boundary at $r=0.71$~Mm is evident in this plot and it is clear that the Gaussian shape of the $v_{\varphi}$ component behaves as expected. The maximum amplitude which the flow obtains is around $4$~km s$^{-1}$. It is worth noting that the rotational flow introduced is torsional in nature, i.e. $\partial/\partial \varphi = 0$, and that the flow is strictly polarised in the azimuthal direction.

\begin{figure*}
\centering
\begin{subfigure}{0.49\textwidth}
    \includegraphics[width=\textwidth]{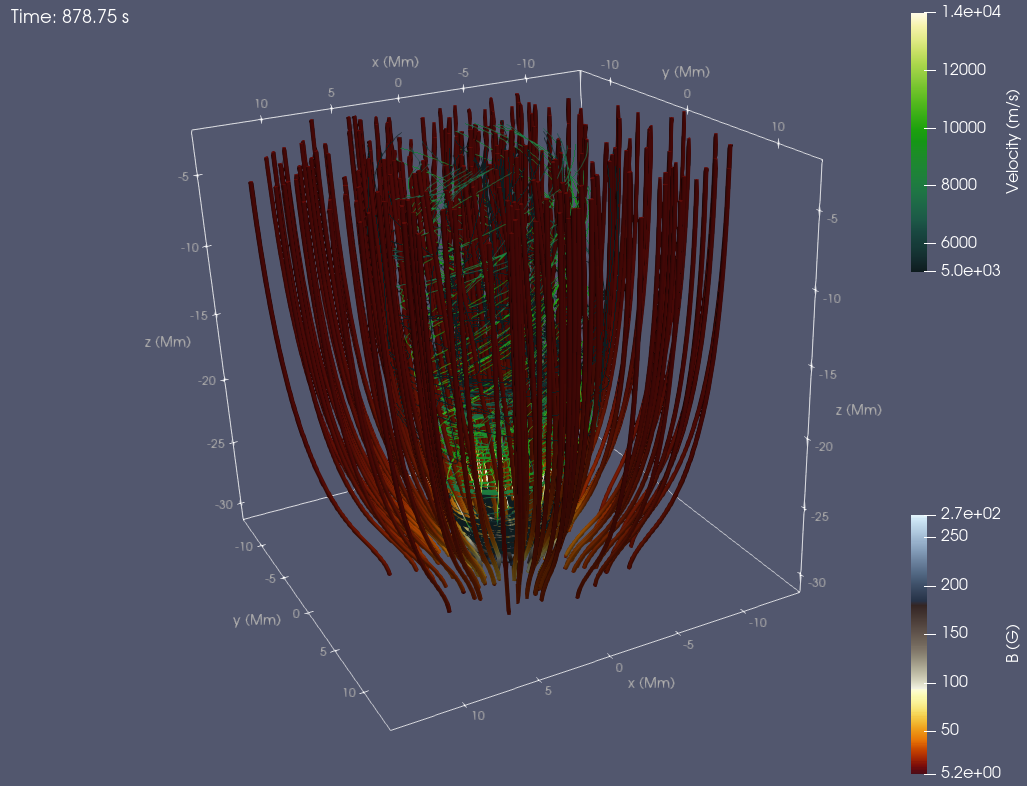}
    \caption{}
    \label{fig:3dsnapshotmagfield}
\end{subfigure}
\begin{subfigure}{0.49\textwidth}
    \includegraphics[width=\textwidth]{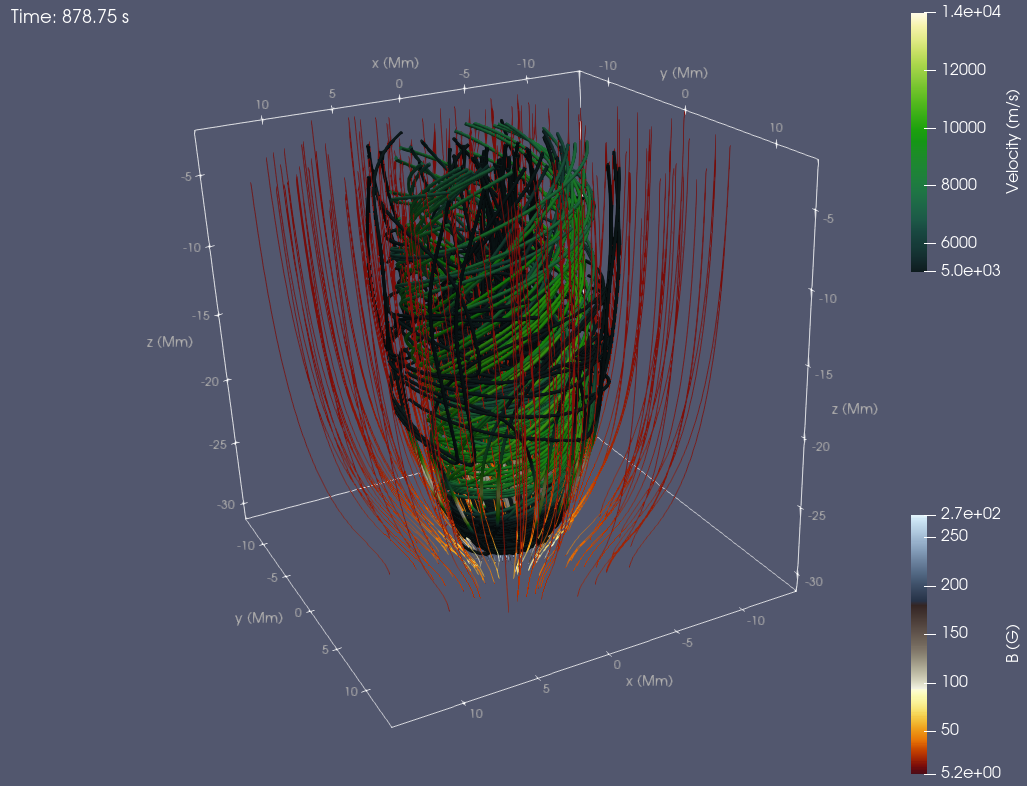}
    \caption{}
    \label{fig:3dsnapshotvelfield}
\end{subfigure}
    
\caption{Snapshot of the 3D streamlines at $t=879$ s of the magnetic field (predominantly red colour) and velocity field (predominantly green colour) for the generated vortex tube in a box outlining the numerical domain from the bottom boundary to the apex. Panel (a) shows the magnetic field as a 3D tube whereas panel (b) highlights the velocity field as a 3D tube.}
\label{fig:3Dstreamlines}
\end{figure*}

A snapshot of the full 3D magnetic and velocity streamlines are displayed in Figure \ref{fig:3dsnapshotmagfield} and Figure \ref{fig:3dsnapshotvelfield}, respectively. This figure highlights the structure of both magnetic and velocity fields resulting from the $v_{\varphi}$ flow applied at the bottom boundary. As the magnetic Reynolds number does not get large enough in the simulation domain for the magnetic field to be advected with the flow, the field lines remain relatively straight and untwisted, however, very small twist can be observed at the bottom boundary where the flow is introduced. With height, the plasma enters the $\beta \ll 1$ regime and the magnetic field ultimately dominates the dynamics of the plasma, resulting in a relatively straight magnetic field in a rotational background plasma flow following a helical structure resulting from a combination of the vertical magnetic field and the $v_{\varphi}$ component of the flow. The strength of the flow is too weak in the corona to influence the magnetic field and creates a rotating structure in a straight magnetic field, properties which are reminiscent of solar vortex tubes appearing naturally in magnetoconvection simulations \citep{Wedemeyer2014, Silva2020}. This is particularly important to emphasise, because it is traditional to assume that the coronal magnetic field is twisted when traced back to the photospheric foot-points where a vortical flow is present in observations. However, it is clear in this simulation that, even when a rotational flow is applied in the photosphere, the magnetic field remains largely untwisted in the corona, whereas it is the plasma velocity which displays a stronger signature of a twisted nature.

\begin{figure}
    \centering
    \includegraphics[width=0.49\linewidth]{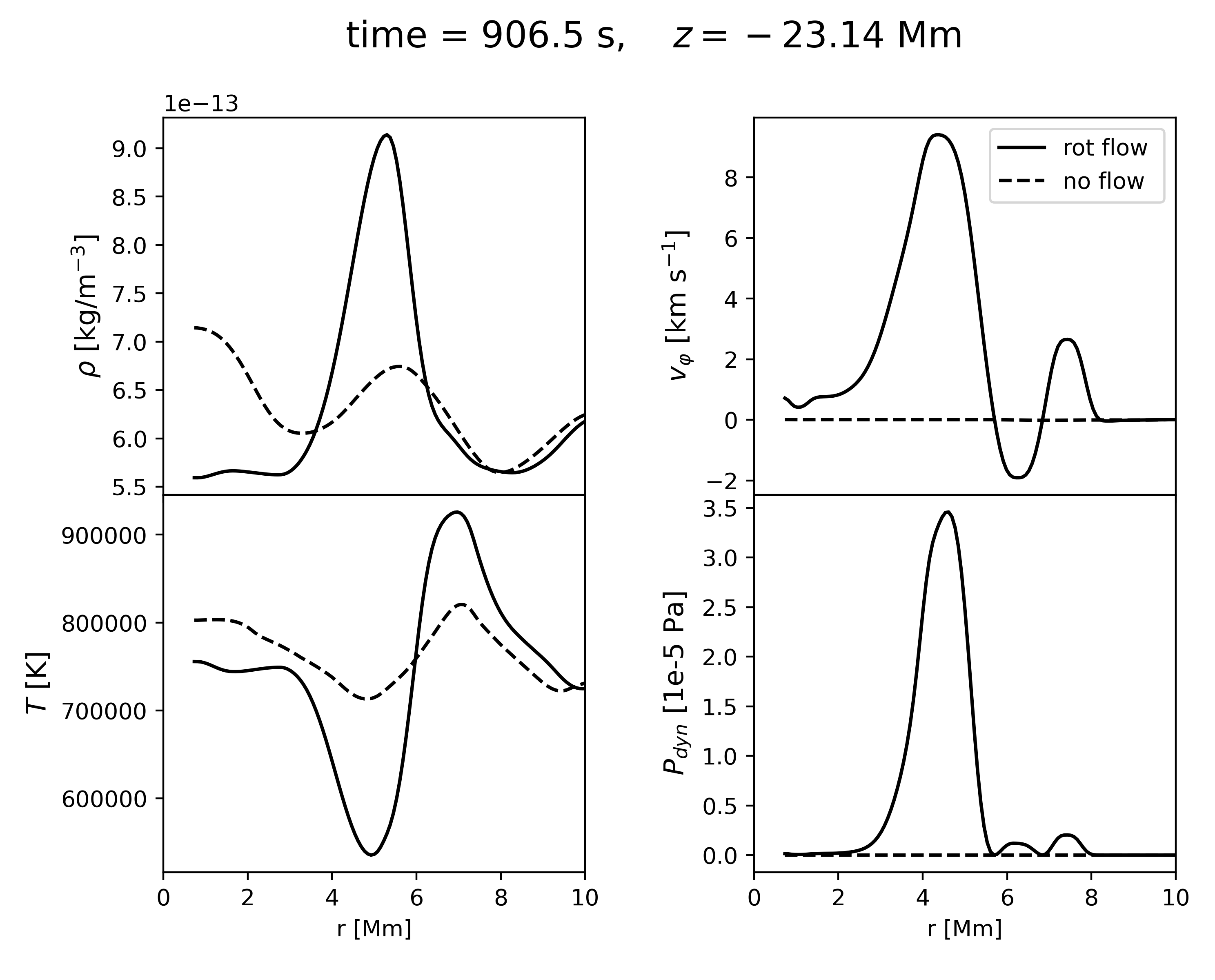}
    \caption{Transverse profiles of the plasma density (top left), rotational component of the velocity (top right), plasma temperature (bottom left) and dynamical pressure (bottom right) in the corona at $t=906.5$~s for simulations with (solid line) and without (dashed line) a rotational flow. These plots are for the simulations excluding a photospheric wave driver, i.e. only consider the effect of the rotational flow.}
    \label{fig:transverse_profiles}
\end{figure}

Figure \ref{fig:transverse_profiles} displays the transverse structuring of the plasma density, temperature, rotational flow and the dynamical pressure for the simulations with and without the rotational flow (excluding the wave driver). The presence of a rotational flow  creates a layer of denser and cooler plasma localised within a thin region (1-2 Mm) where the rotational flow reaches a maximum at the edge of the vortex. This can be understood by considering the effect of dynamical pressure $P_{dyn}$. The dynamical pressure arises from the conservation of fluid momentum in a moving fluid and is defined as $P_{dyn} = v_{\varphi}^2\rho/2$. The dynamical pressure compresses the plasma where the rotational flow is strong, resulting in a layer of cool dense plasma which may be considered as an enhanced waveguide and will be discussed later in terms of supporting wave propagation. Throughout this study we will compare the simulations when the rotational flow is included opposed to when there is no rotational flow present, however, it should be noted that the presence of a rotational flow alters the plasma conditions, so care should be taken to make a complete direct comparison.

\subsection{Wave Driver}\label{subsec:driver}

In addition to the rotational flow outlined in Section \ref{subsec:vortex_formation}, we also incorporate a  gravity-acoustic wave driver which perturbs all components of the velocity vector as well as the plasma pressure and density taking the form \citep{Mihalis1984, Khomenko2012, Santamaria2015, Riedl2021, Skirvin2023ApJ}:
\begin{eqnarray}
    \hat{v}_r = &A& |v| \exp \left(\frac{z}{2H}+\Im(k_z)z\right)\cos(\varphi) \nonumber \\ &\times&\sin \left( \omega t - \Re(k_z)z -k_{\perp}r \cos(\varphi)+\phi_v \right), \label{eq:driver_velocityr}\\    
    \hat{v}_{\varphi}= &-A& |v| \exp \left(\frac{z}{2H}+\Im(k_z)z\right)\sin(\varphi) \nonumber \\ &\times&\sin \left( \omega t - \Re(k_z)z -k_{\perp}r \cos(\varphi)+\phi_v \right), \label{eq:driver_velocityphi}\\            
    \hat{v}_z= &A& \exp \left(\frac{z}{2H}+\Im(k_z)z\right) \nonumber \\ &\times&\sin \left( \omega t - \Re(k_z)z -k_{\perp}r \cos(\varphi)\right), \label{eq:driver_velocityz}\\
    \hat{p} = &A& |P| \exp \left(\frac{z}{2H}+\Im(k_z)z\right) \nonumber \\ &\times&\sin \left( \omega t - \Re(k_z)z -k_{\perp}r \cos(\varphi) + \phi_P\right), \label{eq:driver_pressure}\\
    \hat{\rho} = &A& |\rho| \exp \left(\frac{z}{2H}+\Im(k_z)z\right) \nonumber \\ &\times&\sin \left( \omega t - \Re(k_z)z -k_{\perp}r \cos(\varphi) + \phi_{\rho}\right), \label{eq:driver_density}
\end{eqnarray}
where,
\begin{eqnarray}
    |v| &=& \frac{c_s^2 k_{\perp}}{\left( \omega^2 - c_s^2 k_{\perp}^2\right)} \sqrt{\Re(k_z)^2 + \left(\Im(k_z) + \frac{\gamma - 2}{2H\gamma} \right)^2}, \label{eq:v_amp}\\  
    |P| &=& \frac{\gamma \omega}{\left( \omega^2 - c_s^2 k_{\perp}^2\right)} \sqrt{\Re(k_z)^2 + \left(\Im(k_z) + \frac{\gamma - 2}{2H\gamma} \right)^2}, \label{eq:P_amp}\\  
    |\rho| &=& \frac{\omega}{\left( \omega^2 - c_s^2 k_{\perp}^2\right)} \nonumber \\ &\times& \sqrt{\Re(k_z)^2 + \left(\Im(k_z) - \frac{1}{2H} + \frac{\gamma - 1}{H\gamma}\frac{k_{\perp}^2c_s^2}{\omega^2} \right)^2}, \label{eq:rho_amp}\\       
     k_r &=& \cos(\varphi) \sin(\theta)\left|\frac{\omega}{c_s}\sqrt{\frac{(\omega_c^2 -\omega^2)}{(\omega_g^2 \sin(\theta)^2-\omega^2)}}\right|, \label{eq:kr}\\   
     k_{\varphi} &=& \sin(\varphi) \sin(\theta)\left|\frac{\omega}{c_s}\sqrt{\frac{(\omega_c^2 -\omega^2)}{(\omega_g^2 \sin(\theta)^2-\omega^2)}}\right|, \label{eq:kphi}\\   
     k_{\perp} &=& \sqrt{k_r^2 + k_{\varphi}^2}, \label{eq:kperp}\\ 
     k_z &=& \sqrt{\frac{\omega^2 - \omega_c^2}{c_s^2} - k_{\perp}^2\frac{\omega^2-\omega_g^2}{\omega^2}}, \label{eq:kz}  \\
     \omega_c &=& \frac{c_s}{2H}\cos(\theta_B),
     \label{eq:omega_c} \\   
     \omega_g &=& \frac{2\omega_c\sqrt{\gamma-1}}{\gamma}, \label{eq:omega_g} \\
     \phi_P &=& \arctan\left(\frac{\Im(k_z) + (\gamma-2)/2 H\gamma}{\Re(k_z)}\right), \label{eq:phi_p} \\
     \phi_v &=& \phi_P, \label{eq:phi_v} \\
     \phi_{\rho} &=& \arctan\left(\frac{\Im(k_z) -\frac{1}{2H} + \frac{(\gamma-1)(c_s^2 k_{\perp}^2/\omega^2)}{\gamma H}}{\Re(k_z)}\right). \label{eq:phi_rho}
\end{eqnarray}
Here, $A=300$ m s$^{-1}$ is the driver amplitude, which agrees with photospheric Doppler oscillations from the contribution of p-modes \citep{McClure2019}. This amplitude is also weaker than the rotational flow applied at the foot-point, which also increases with time (see Figure \ref{fig:vphi_with_time}). The amplitudes for the velocity, pressure and density perturbations are denoted as $|v|$, $|P|$ and $|\rho|$, respectively; $H$ is the pressure scale height; $\gamma=5/3$ is the adiabatic index for a monatomic gas, $k_{\perp}$ is the horizontal wavenumber of the driven waves and $k_z$ is the vertical wavenumber, which only has a real part in our case; $\omega = 2\pi/T$ is the driver frequency, with period $T = 370$ s, which is within the typical range of p-mode periods. The variables $\phi_v$, $\phi_P$ and $\phi_{\rho}$ are the velocity, pressure and density phase shifts compared to the vertical velocity perturbation, $\hat{v}_z$. The acoustic cut-off and thermally modified acoustic cut-off frequencies are denoted as $\omega_c$ and $\omega_g$, respectively, where $\theta_B$ is the angle between the magnetic field and direction of gravitational acceleration. The angle of the driver with respect to the vertical axis is represented by $\theta$ and is taken to be $\theta = 15^{\circ}$ in order to excite non-axisymmetric modes \citep{Skirvin2023ApJ, Gao2023}.

The driver is applied locally at only one foot-point at the base of the domain, and located inside the vortex tube. This is achieved by multiplying Equations (\ref{eq:driver_velocityr})-(\ref{eq:driver_density}) by the function:
\begin{equation}
    D(r) = \exp\left( - \frac{r^2}{\sigma^2}\right),
    \label{eq:Gaussian_driver}
\end{equation}
where $\sigma$ is the standard deviation of the Gaussian profile describing the width of the driver, which in this study is taken to be $\sigma = 2$ Mm. At the photospheric base of the domain, the full width half maximum of the Gaussian magnetic field strength is located at $r = 2.58$ Mm, therefore the driver can be considered to be applied within the foot-point of the vortex tube.

In order to investigate MHD wave propagation in the simulated vortex tube, we run two simulations, one with the photospheric wave driver and one without the wave driver. Therefore, by subtracting the simulation without the wave driver from the simulation with the wave driver, we can study the effect of the driven perturbations within the vortex. It is important to note however, that this is not equivalent to a linearisation and that all non-linear effects will still be included. Moreover, we run an additional simulation with the inclusion of the photospheric wave driver but without the additional rotational flow described in Section \ref{subsec:vortex_formation}, this will allow us to compare the properties of MHD waves in the presence of a rotational plasma flow to a setup without such a background flow.

\section{Results} \label{sec:results}

\subsection{Wave Decomposition}\label{subsec:wavedecomp}

In full 3D simulations, where the magnetic field is not confined to a 2D geometry, the isolation of MHD waves becomes non-trivial as there are an infinite number of vectors normal to the magnetic field vector \citep{Yadav2022}. To help distinguish between the different types of waves in our simulation, we decompose motions relative to circular contours of constant magnetic field strength \citep{Riedl2019, Skirvin2024modeconv}. The conversion of components from a cylindrical geometry ($r, \varphi, z$) to those parallel, perpendicular and azimuthal to magnetic flux surfaces is given by:
\begin{eqnarray}
    \mathbf{e}_{\parallel} &=& \left[\frac{B_r \text{cos}(\varphi)}{\sqrt{B_r^2 + B_z^2}}, \frac{B_r \text{sin}(\varphi)}{\sqrt{B_r^2 + B_z^2}}, \frac{B_z(\varphi)} {\sqrt{B_r^2 + B_z^2}}\right], \label{decomp_parallel} \\
    \mathbf{e}_{\phi} &=& [-\text{sin}(\varphi), \text{cos}(\varphi), 0], \label{decomp_azi} \\
    \bf{e}_{\perp} &=& \bf{e}_{\phi} \times \bf{e}_{\parallel}, \label{decomp_perp}
\end{eqnarray}
where $\mathbf{e}$ denotes a unit vector in each direction, respectively. Equations (\ref{decomp_parallel})-(\ref{decomp_perp}) set up a Cartesian basis describing the vector decomposition with respect to magnetic field lines for an equilibrium magnetic field which is structured in the $r$ and $z$ directions in a cylindrical geometry. The component of magnetic field azimuthal to magnetic surfaces is ignored in the decomposition due to the field lines being circularly symmetric around the axis of the vortex tube and uniform rotational velocity is considered in the initial model (independent of $\varphi$). Moreover, the pitch angle of the magnetic field remains very small ($B_z \gg B_{\varphi}$; see Figure \ref{fig:3Dstreamlines}), especially in the plasma-$\beta \ll 1$ regime. This decomposition of components parallel, perpendicular and azimuthal to the magnetic field lines will be important in the context of understanding the wave modes which are present in the simulation.

\begin{figure*}[ht!]
\centering
\includegraphics[width=0.95\textwidth]{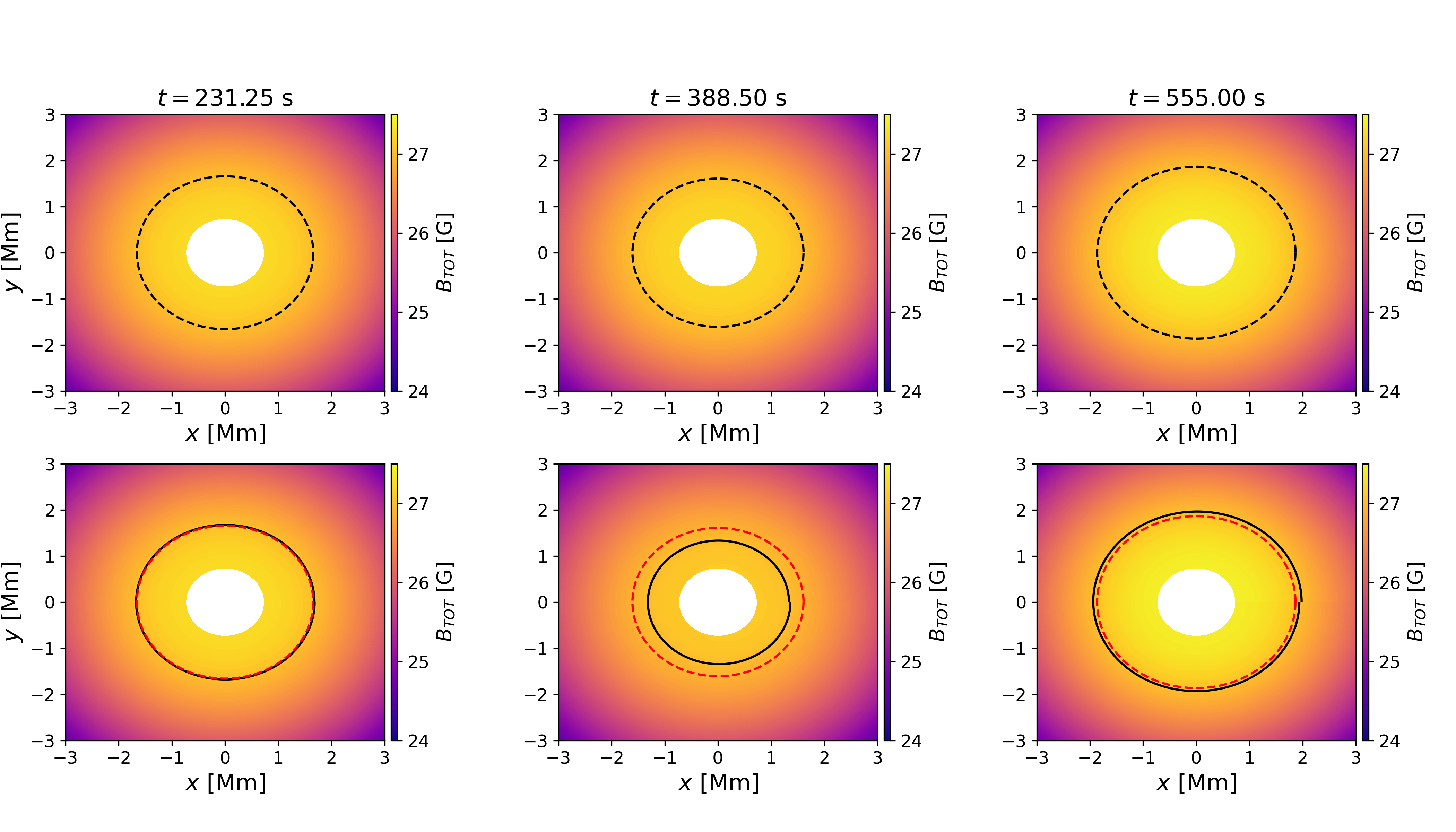}
\caption{2D contours of the magnetic field strength in the corona at a height of $z=-23.14$ Mm. Three different time snapshots are shown in different columns. The top row shows the time snapshots from the simulation including a rotational flow but with no wave driver, whereas the bottom row highlights the simulation with a rotational flow and a photospheric wave driver. The black contours indicate a constant magnetic field contour of $27$ G. The red dashed contours in the bottom panels indicate the same magnetic contour for the simulation without a wave driver (i.e. the contours shown in the top panel).}
\label{fig:corona_magfield_contours}
\end{figure*}

It should be noted here the validity of the decomposition described by Equations (\ref{decomp_parallel})-(\ref{decomp_perp}) in the presence of a background rotational flow or a weakly twisted magnetic field. As the $v_{\varphi}$ perturbation applied at the lower boundary is uniform in the azimuthal direction, and the wave driver amplitude is smaller than the background flow, it is suitable to assume that the contours of constant magnetic field remain circular in a two-dimensional cross-cut. This is highlighted also in Figure \ref{fig:corona_magfield_contours}, where it is clear that, even in the presence of a background rotational flow, the constant magnetic field contours retain their circular outline throughout the duration of the simulation, such that Equations (\ref{decomp_parallel})-(\ref{decomp_perp}) remain valid for the present analysis. Therefore, any discussion regarding wave mode decomposition in this study will focus on the corona where the magnetic field is untwisted ($B_z \gg B_{\varphi}$). Figure \ref{fig:corona_magfield_contours} highlights that the magnetic field naturally expands in the corona with time, even in the absence of a photospheric wave driver, due to the rotational flow creating a radial pressure gradient which causes the flux tube structure to slightly expand with time.

\subsection{Velocity perturbations}\label{subsec:velpert}

The magnetic contours in Figure \ref{fig:corona_magfield_contours} demonstrate the presence of the sausage mode within the vortex tube. This is evident in the bottom panels of Figure \ref{fig:corona_magfield_contours} through expanding and contracting of the contour of constant magnetic flux, reminiscent of a sausage mode in a cylindrical flux tube. The wave driver adopted in the current study is known to excite sausage modes \citep{Riedl2021, Skirvin2023ApJ}, moreover vortex motions have been shown to excite sausage modes in flux tubes even in the absence of a wave driver \citep{fedshe2011}, therefore, this analysis provides further evidence of sausage modes supported by vortex tubes in the solar corona. Furthermore, the asymmetric wave driver also produces kink modes \citep{Riedl2019, Skirvin2023ApJ, Gao2023}, however, due to the vertical component of the perturbed velocity being considerably stronger than the horizontal components of the perturbed velocity, it is difficult to appreciate the horizontal displacement of the magnetic contours in Figure~\ref{fig:corona_magfield_contours}.

\begin{figure*}
\centering
\includegraphics[width=0.95\textwidth]{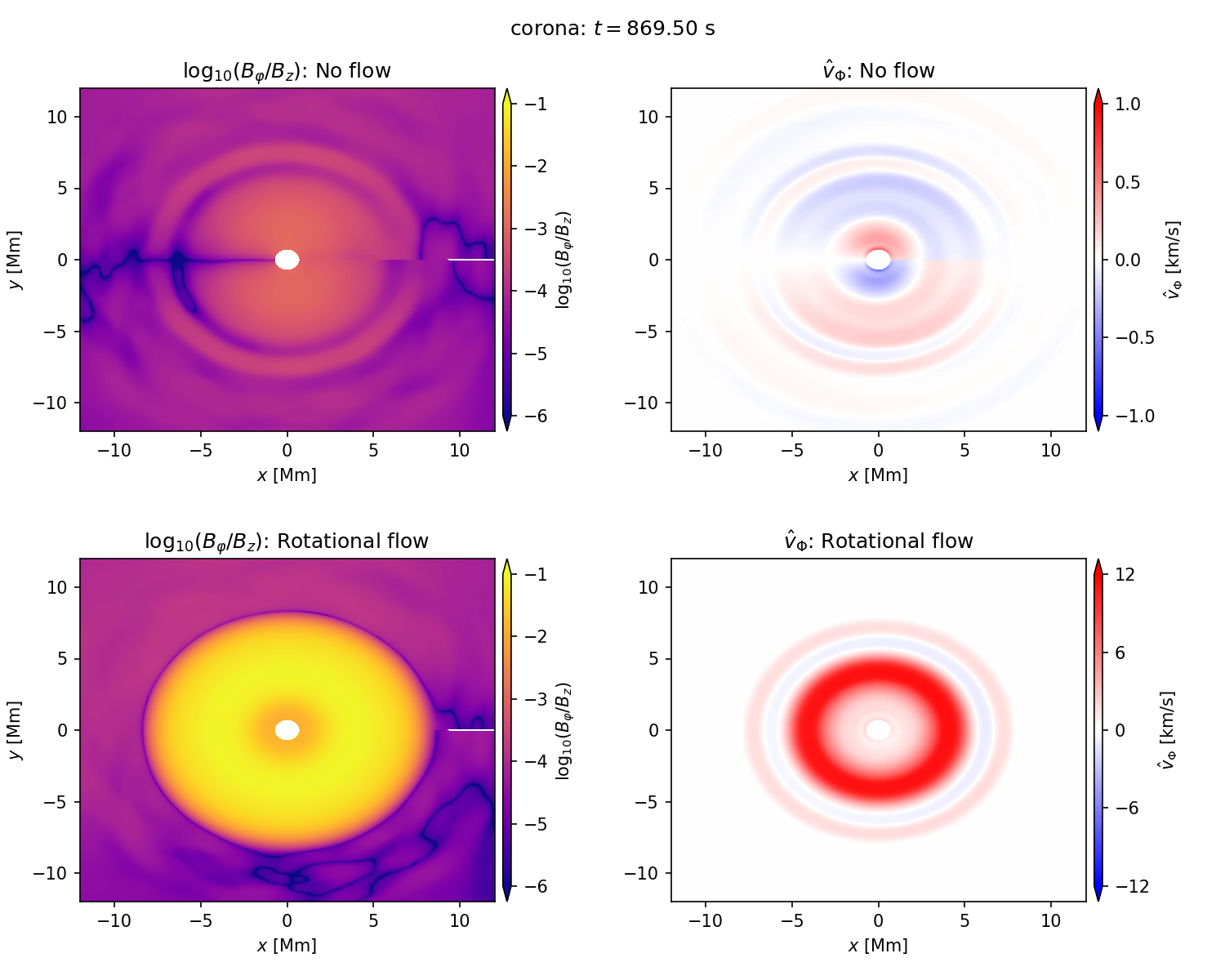}
\caption{Horizontal cross-cuts at a coronal height ($z=-23.14$ Mm) displaying the logarithm of the ratio of the total azimuthal component of the magnetic field, $B_{\varphi}$, to the total vertical component of the magnetic field, $B_z$, as a proxy for the pitch angle of the magnetic field. In addition, we show the perturbed velocity tangential to constant magnetic contours $\hat{v}_{\phi}$ as defined in Equation (\ref{decomp_azi}). The top row shows the simulation with no rotational perturbation applied, whereas the bottom row indicates the case with a rotational flow at one foot-point. It should be noted that the colourbar is saturated for the magnetic field (left hand panels).}
\label{fig:corona_decomp}
\end{figure*}
Using the decomposition method outlined in Section \ref{subsec:wavedecomp}, we analyse the velocity perturbations tangential to circular contours of constant magnetic field strength. Figure \ref{fig:corona_decomp} displays the total magnetic twist (log$_{10}(B_{\varphi}/B_z)$) and the $\hat{v}_{\phi}$ perturbation decomposition at a coronal height ($z = -23.14$~Mm) for the simulation without the rotational flow and the simulation with the rotational flow incorporated. It is important to note here that the $B_{\varphi}$ component is still an order of magnitude weaker than the $B_z$ component at this coronal height for the simulation with a rotational flow, further validating the decomposition outlined in \ref{subsec:wavedecomp}. Firstly, let us concentrate on the simulation without a rotational flow added. The nature of the wave driver described in Section \ref{subsec:driver}, generates transverse (kink) waves inside the vortex tube \citep{Riedl2019, Skirvin2023ApJ, Gao2023}. When these motions are decomposed with respect to the magnetic flux surfaces outlined in Section \ref{subsec:wavedecomp}, the clear asymmetric pattern of positive and negative fluctuations for $y>0$~Mm and $y<0$~Mm highlights the nature of the linearly polarised kink mode (top right panel of Figure \ref{fig:corona_decomp}). This is further understood when looking at Figure 3 in the study by \citet{goo2014} highlighting the rotational component of kink modes. The $B_{\varphi}$ in this case is purely a result of the kink mode perturbation and is much weaker than the ambient total magnetic field. However, for the simulation with the addition of a rotational plasma flow, the total magnetic twist is stronger and directed in one orientation. The driven waves in the rotational plasma flow, when decomposed tangential to flux surfaces, display a very different pattern to the scenario of a flux tube in the absence of rotational flow. The bottom right panel of Figure \ref{fig:corona_decomp} shows the $\hat{v}_{\phi}$ decomposition of the motions in the simulation with a rotational flow and a wave driver. The resulting perturbations closely mimic those expected from torsional Alfv\'{e}n waves, in the sense that they exhibit periodic twisting motions which are azimuthally symmetric (also see Figure 3 of \citet{goo2014}). However, it is naive to interpret these motions as an indication to the presence of torsional Alfv\'{e}n waves in the simulation as, any torsional Alfv\'{e}n waves generated by the non-stationary rotational flow would still produce one sign of the velocity perturbation (i.e. either positive or negative) instead of an oscillatory signal, however, it is evident that there is an oscillatory signal present which suggests these motions arise from the wave driver implemented at the bottom boundary. Instead, these perturbations more likely hint to the presence of sausage modes present within the vortex tube. As discussed earlier, the wave driver is known to produce sausage modes within the vortex tube \citep{Riedl2021}, however, it has been shown that sausage modes in twisted magnetic flux tubes can also produce azimuthal perturbations \citep{Giag_2015} with similar signatures to torsional Alfv\'{e}n waves.

\subsection{Azimuthal wavenumbers}\label{subsec:aziwavenumbers}

To investigate the MHD modes supported by the simulated vortex tube, we can apply a local Fourier decomposition to quantify the contribution of different azimuthal wavenumbers in our simulation. We employ a discrete Fourier transform to analyse the contribution of the different azimuthal wavenumbers using the profiles of $\hat{v}_r$ and $\hat{v}_{\varphi}$ 
 as a function of angle at a fixed radial location \citep{ter2018, magyar2022, Skirvin2023ApJ}. The notation $m$ is adopted to represent the excited azimuthal wavenumbers in the simulation, and should not be confused with the eigenmode solutions. It is then possible to write the discrete Fourier transform, namely, the function $g$, on each flux surface as:
\begin{equation}
    G(m) = \frac{1}{N} \sum_{k=0}^{N-1} g(k) e^{-i\frac{2\pi}{N}mk},
\end{equation}
for a discrete set of N samples ($m=0, ..., N-1$). In our case, the analysis is in the azimuthal direction, ranging from $0$ to $2\pi$. This means that instead of using $k$, it is more convenient to introduce the parameter $\theta_k = 2\pi k/N$. The contribution of each excited wave mode $p$ to the total signal can be expressed, using the inverse Fourier transform, as:
\begin{equation}
    g(\theta_k) = \sum_{m=0}^{N-1} G(m) e^{im\theta_k}.
\end{equation}
Unlike previous analyses \citep{ter2018, magyar2022, Skirvin2023ApJ}, the spectrum of excited azimuthal modes is no longer symmetric about $N/2$, hence, the Fourier transform of the signals $\hat{v}_r$ and $\hat{v}_{\varphi}$ are no longer purely real or imaginary corresponding to cosine and sine components, respectively, for the simulation with a rotational flow applied. This result is easy to understand as the rotational flow breaks the azimuthal symmetry for the perturbed signals, which can no longer be separated into solely sine or cosine functions, instead, they are represented by a combination of even and odd functions which can be modelled by both sine and cosine functions.

\begin{figure*}
\centering
\begin{subfigure}{0.48\textwidth}
    \includegraphics[width=\textwidth]{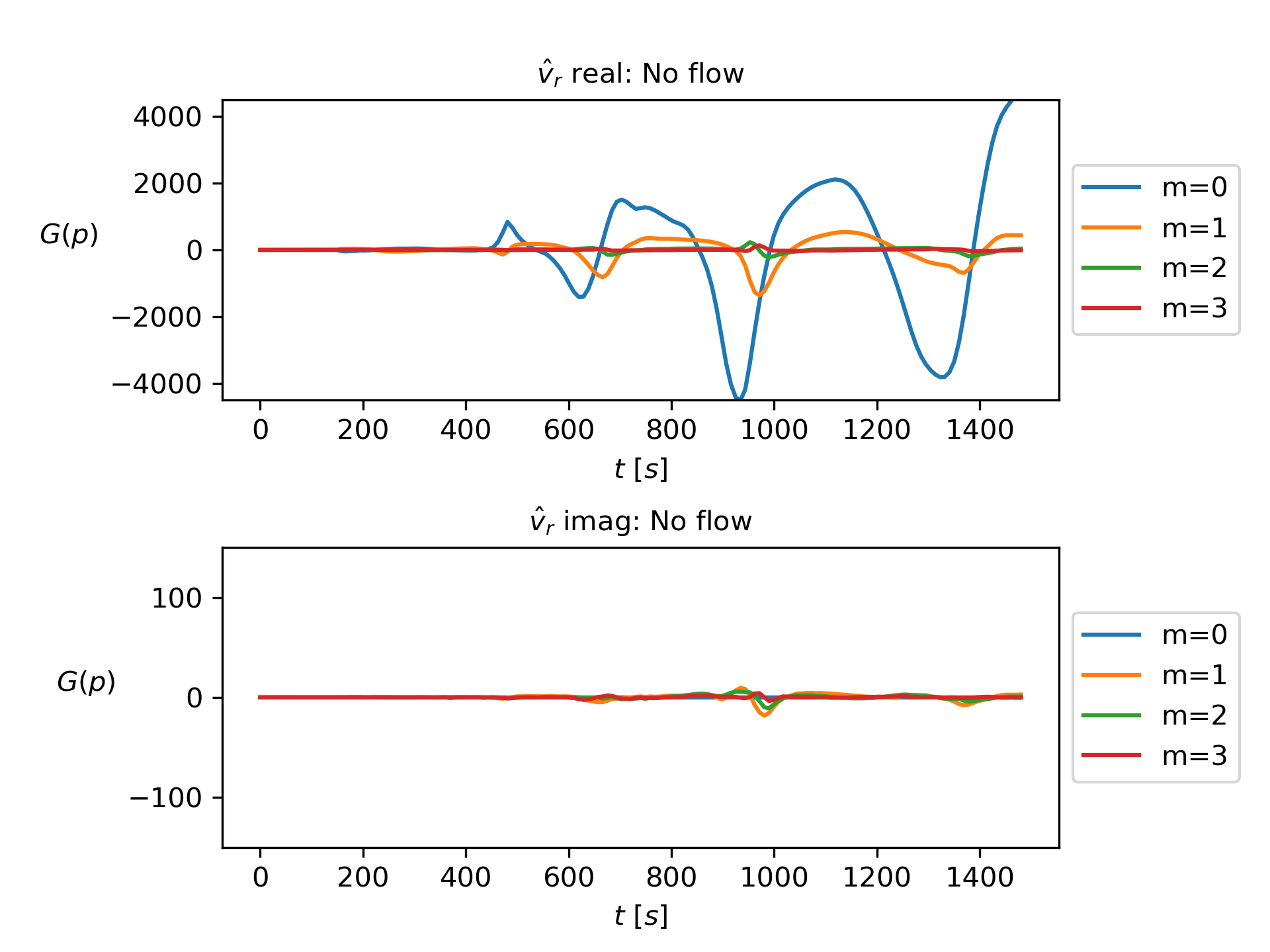}
    \caption{}
    \label{fig:noflow_vr_FFT}
\end{subfigure}
\begin{subfigure}{0.48\textwidth}
    \includegraphics[width=\textwidth]{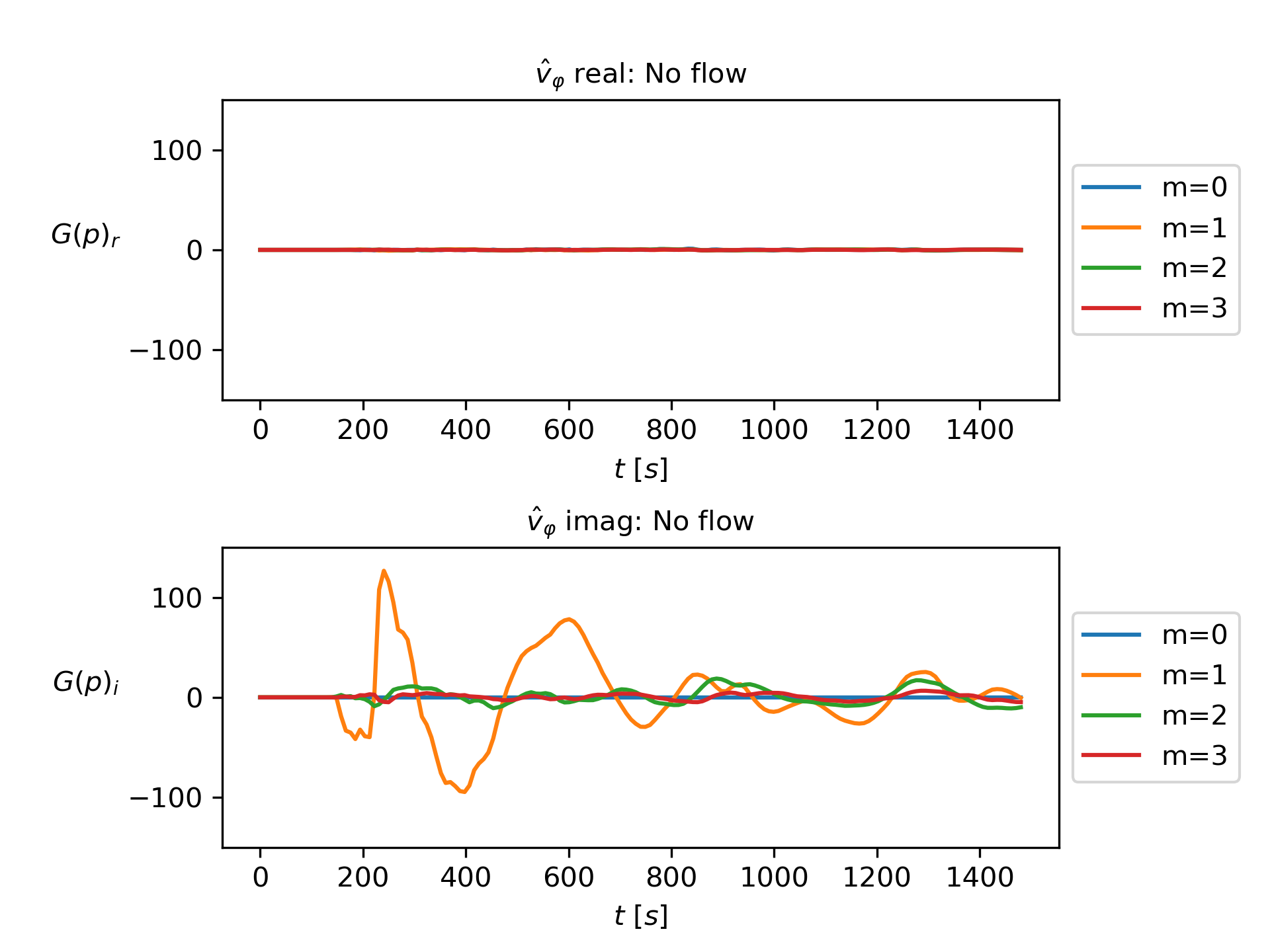}
    \caption{}
    \label{fig:noflow_vphi_FFT}
\end{subfigure}
\begin{subfigure}{0.48\textwidth}
    \includegraphics[width=\textwidth]{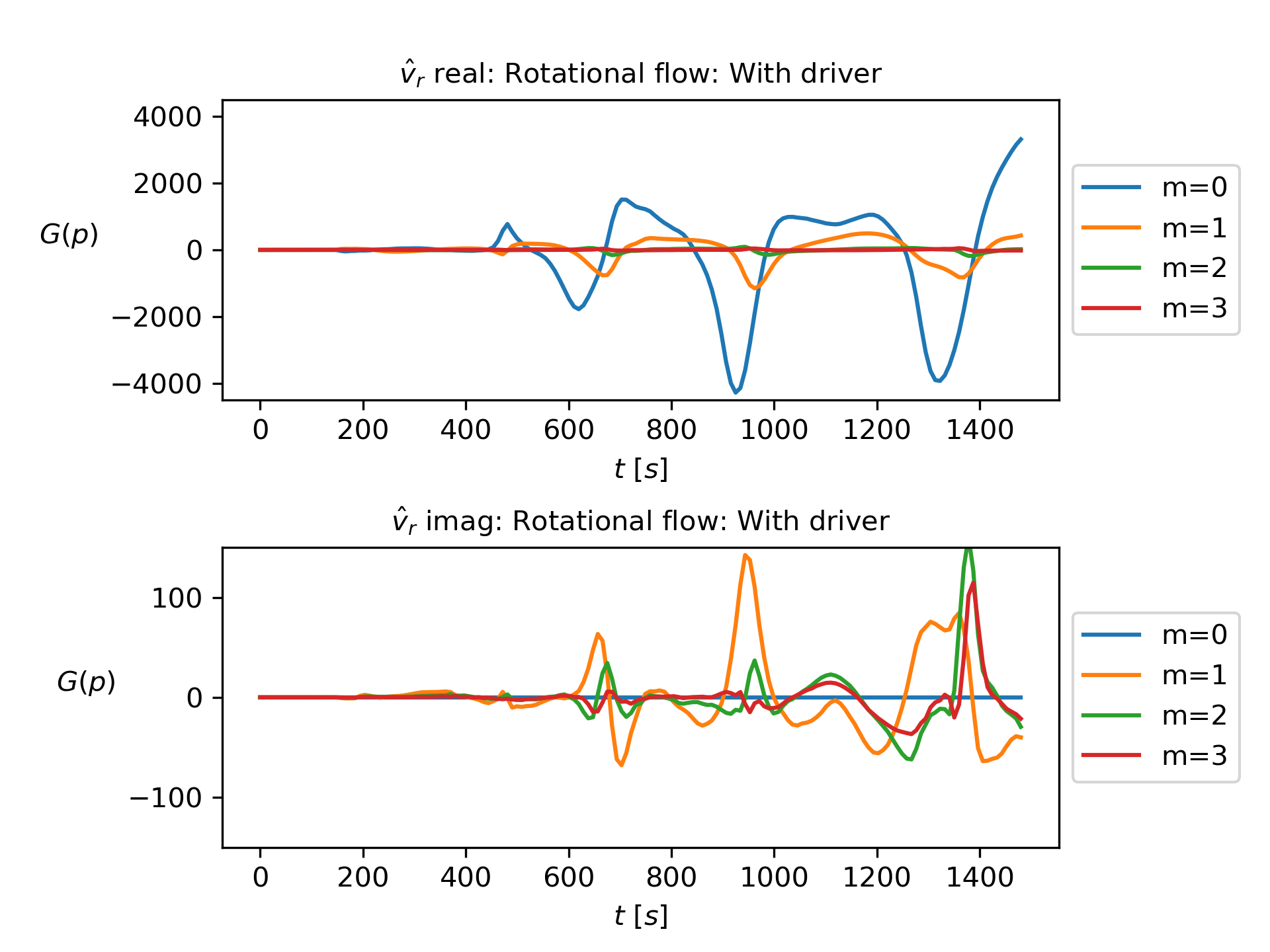}
    \caption{}
    \label{fig:rotflow_vr_FFT}
\end{subfigure}
\begin{subfigure}{0.48\textwidth}
    \includegraphics[width=\textwidth]{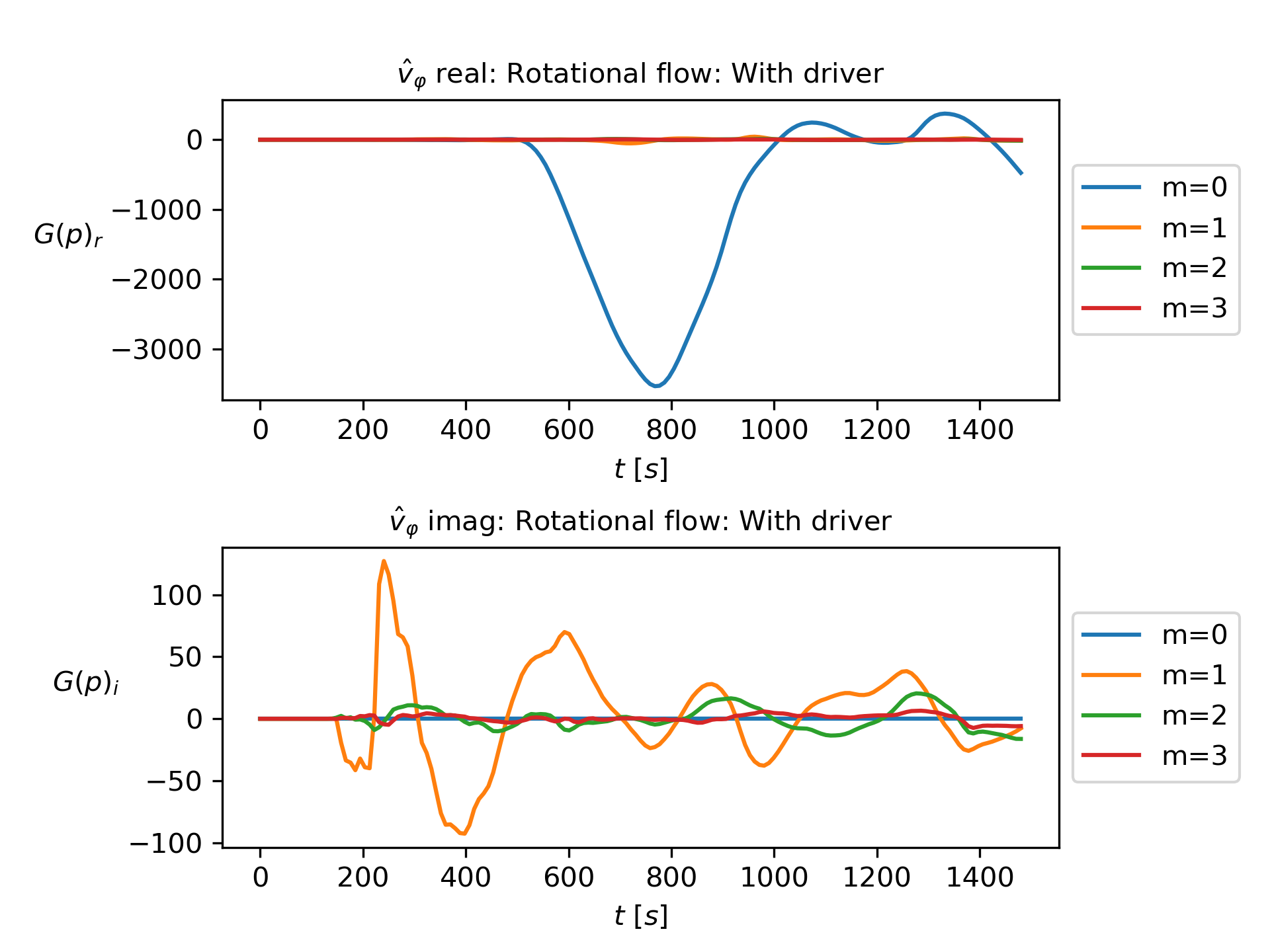}
    \caption{}
    \label{fig:rotflow_vphi_FFT}
\end{subfigure}
\caption{The real and imaginary Fourier coefficients of $\hat{v}_r$ and $\hat{v}_{\varphi}$ in the corona ($z=-23.14$ Mm) around flux surfaces at a radial location of $r=7.20$~Mm. Panels (a) and (b) show the Fourier coefficients for the simulation without a rotational flow (i.e. just a photospheric wave driver) same as \citet{Skirvin2023ApJ}, for $\hat{v}_r$ and $\hat{v}_{\varphi}$, respectively. Panels (c) and (d) show the same components but for the simulation with both a rotational flow and a photospheric wave driver. The different azimuthal wavenumbers excited are represented by different colours, highlighted in the legend.}
\label{fig:fourier_coefficients}
\end{figure*}

The disparity of the resulting Fourier signals is highlighted in Figure \ref{fig:fourier_coefficients}, where the Fourier coefficients for $\hat{v}_r$ and $\hat{v}_{\varphi}$ are displayed for both simulations with and without a rotational flow. It is evident from Figure \ref{fig:noflow_vr_FFT} and Figure \ref{fig:noflow_vphi_FFT} that when no rotational flow is present, the azimuthal signals neatly decompose into even and odd functions for $\hat{v}_r$ and $\hat{v}_{\varphi}$, respectively, highlighting the presence of sausage and kink modes resulting from the wave driver. However, when a rotational flow is present, there is significant power in both the real and imaginary Fourier coefficients as displayed in Figure \ref{fig:rotflow_vr_FFT} and Figure \ref{fig:rotflow_vphi_FFT}. Furthermore, in the presence of a background rotational flow, a significant power is associated with the axisymmetric $m=0$ mode in the decomposed real $\hat{v}_r$ and $\hat{v}_{\varphi}$ signal. The Fourier analysis allows us to distinguish the MHD modes present by comparing the periodicities of these signals. For example, the $m=0$ mode identified in the real $\hat{v}_r$ signal is evidence of the sausage mode resulting from the photospheric wave driver and supported by the vortex tube. This is clear because the period of this signal is equivalent to the period of the higher order modes (e.g. $m=1$ kink mode) which is only generated by the photospheric wave driver. On the other hand, the strong $m=0$ coefficient in the real $\hat{v}_{\varphi}$ signal is evidence of the torsional Alfv\'{e}n wave resulting from the temporal behaviour of the rotational flow applied at the photospheric foot-point in the simulation. This can be seen by comparing the periodicity of this signal with that displayed in Figure \ref{fig:vphi_with_time} and appreciating that these motions will possess a time-lag when observed in the corona as displayed in Figure \ref{fig:rotflow_vphi_FFT}. We can calculate the crossing time required for an Alfv\'{e}n wave to travel to this height given that the Alfv\'{e}n speed below the TR is around $10-15$~km/s and the TR is positioned approximately $6$~Mm above the bottom boundary. This results in a time lag of roughly $400-600$~s, agreeing with the lag observed by comparing Figure \ref{fig:noflow_vphi_FFT} and Figure \ref{fig:rotflow_vphi_FFT}. As the time derivative of the rotational flow is non-zero it generates incompressible torsional oscillations which propagate within the vortex tube as torsional Alfv\'{e}n waves (or pulses). 

\begin{figure*}
\centering
\begin{subfigure}{0.48\textwidth}
    \includegraphics[width=\textwidth]{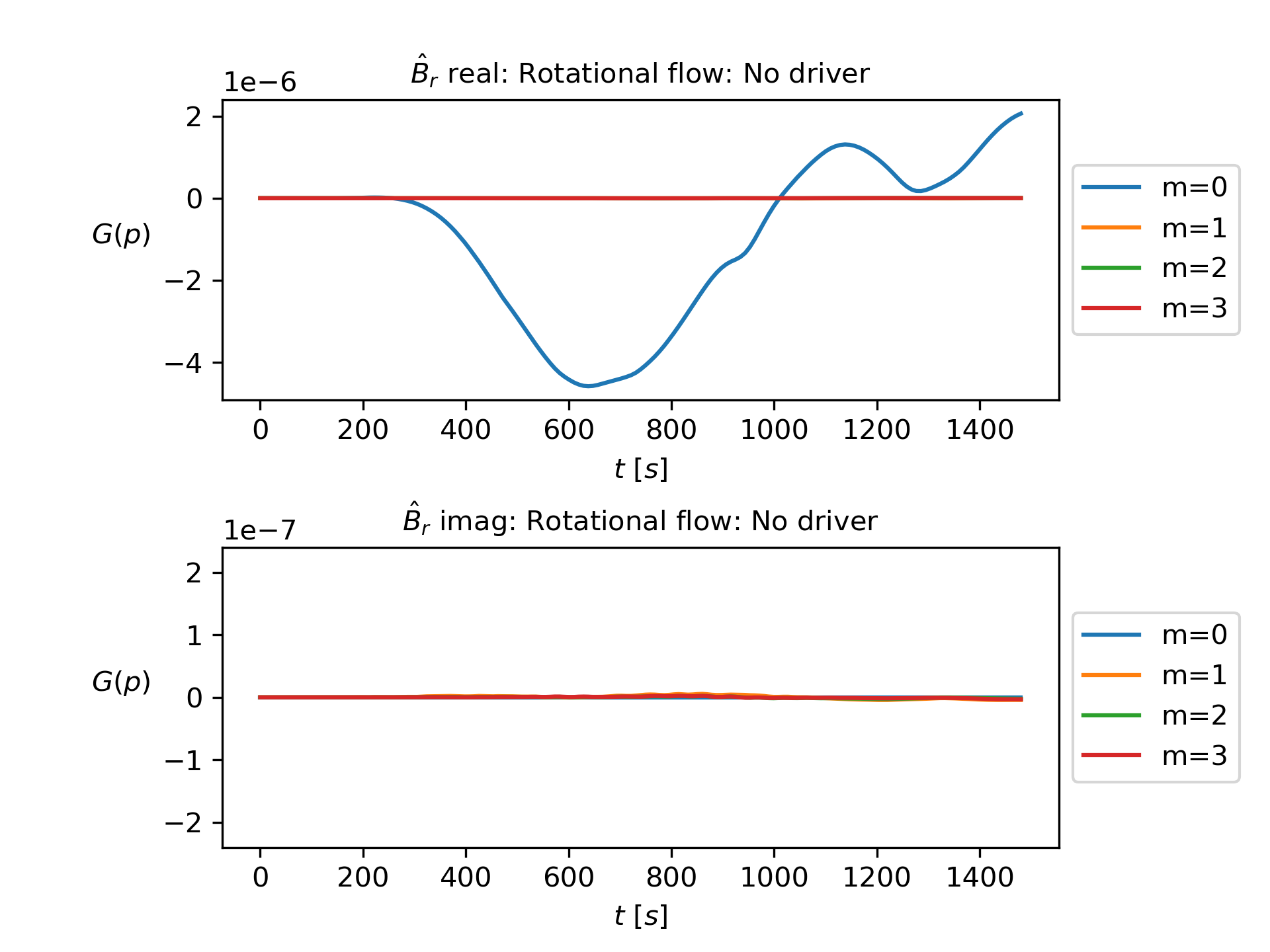}
    \caption{}
    \label{fig:rotflow_nodriver_b_FFT}
\end{subfigure}
\begin{subfigure}{0.48\textwidth}
    \includegraphics[width=\textwidth]{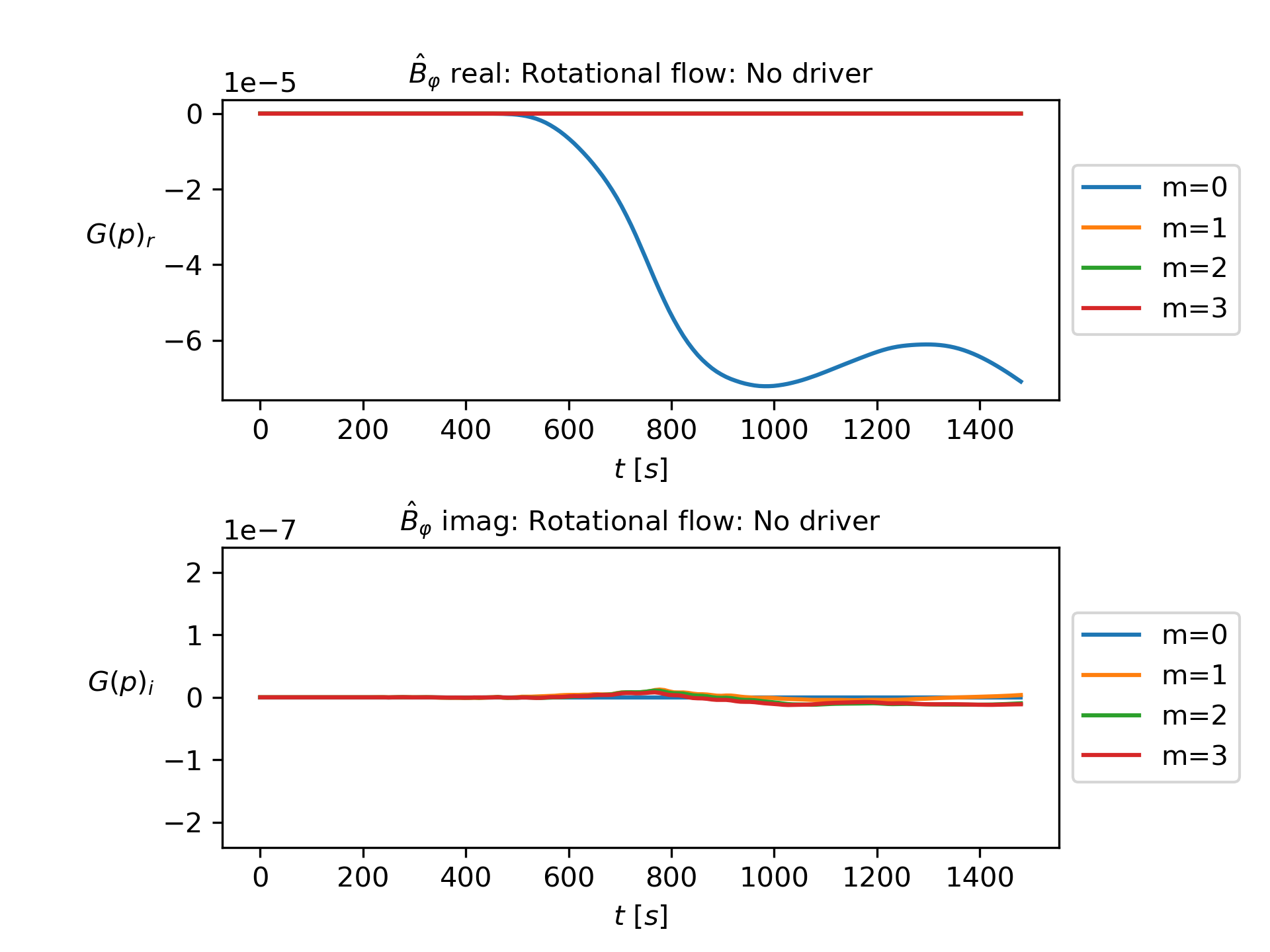}
    \caption{}
    \label{fig:rotflow_nodriver_bphi_FFT}
\end{subfigure}
\begin{subfigure}{0.48\textwidth}
    \includegraphics[width=\textwidth]{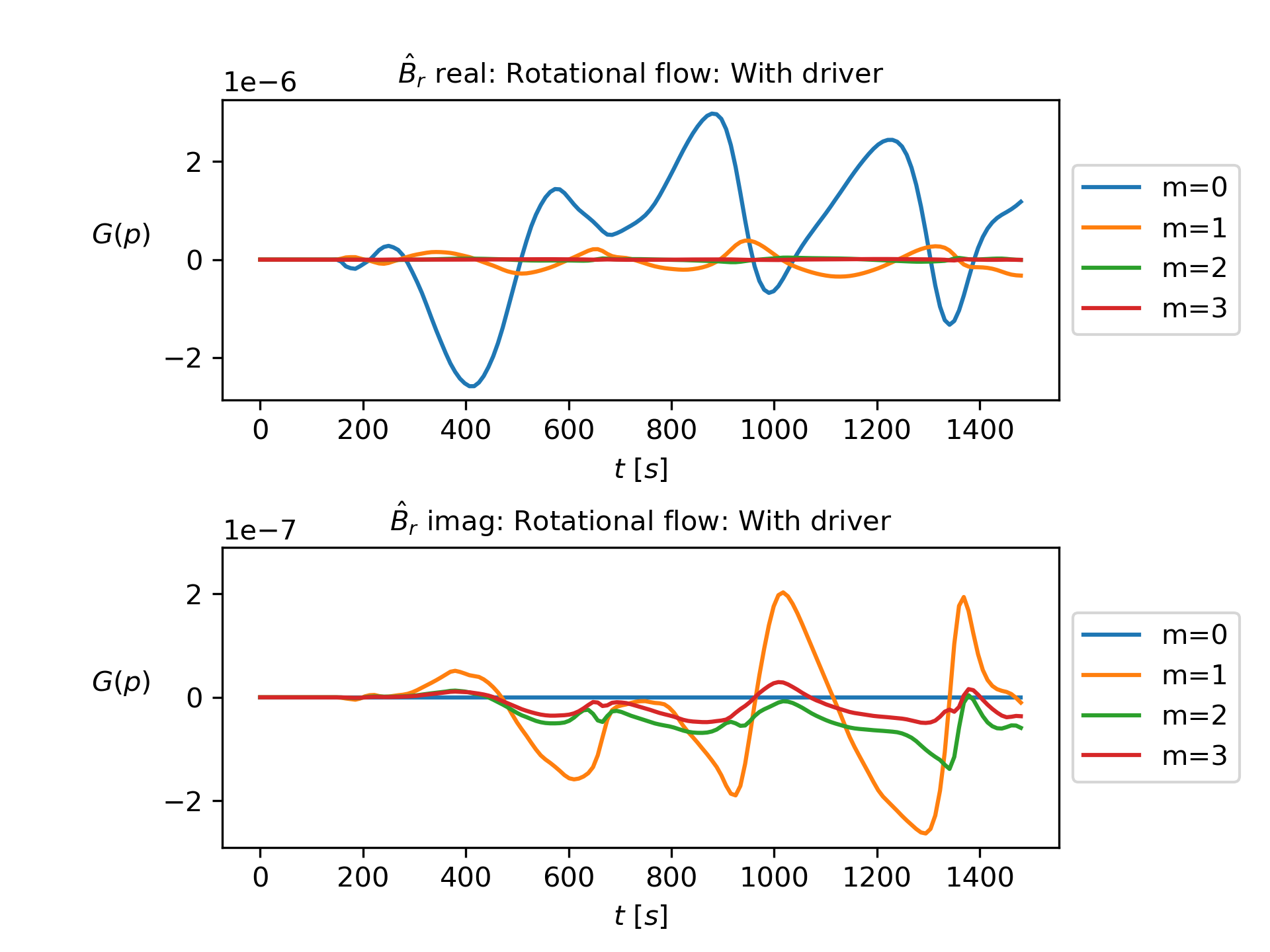}
    \caption{}
    \label{fig:rotflow_br_FFT}
\end{subfigure}
\begin{subfigure}{0.48\textwidth}
    \includegraphics[width=\textwidth]{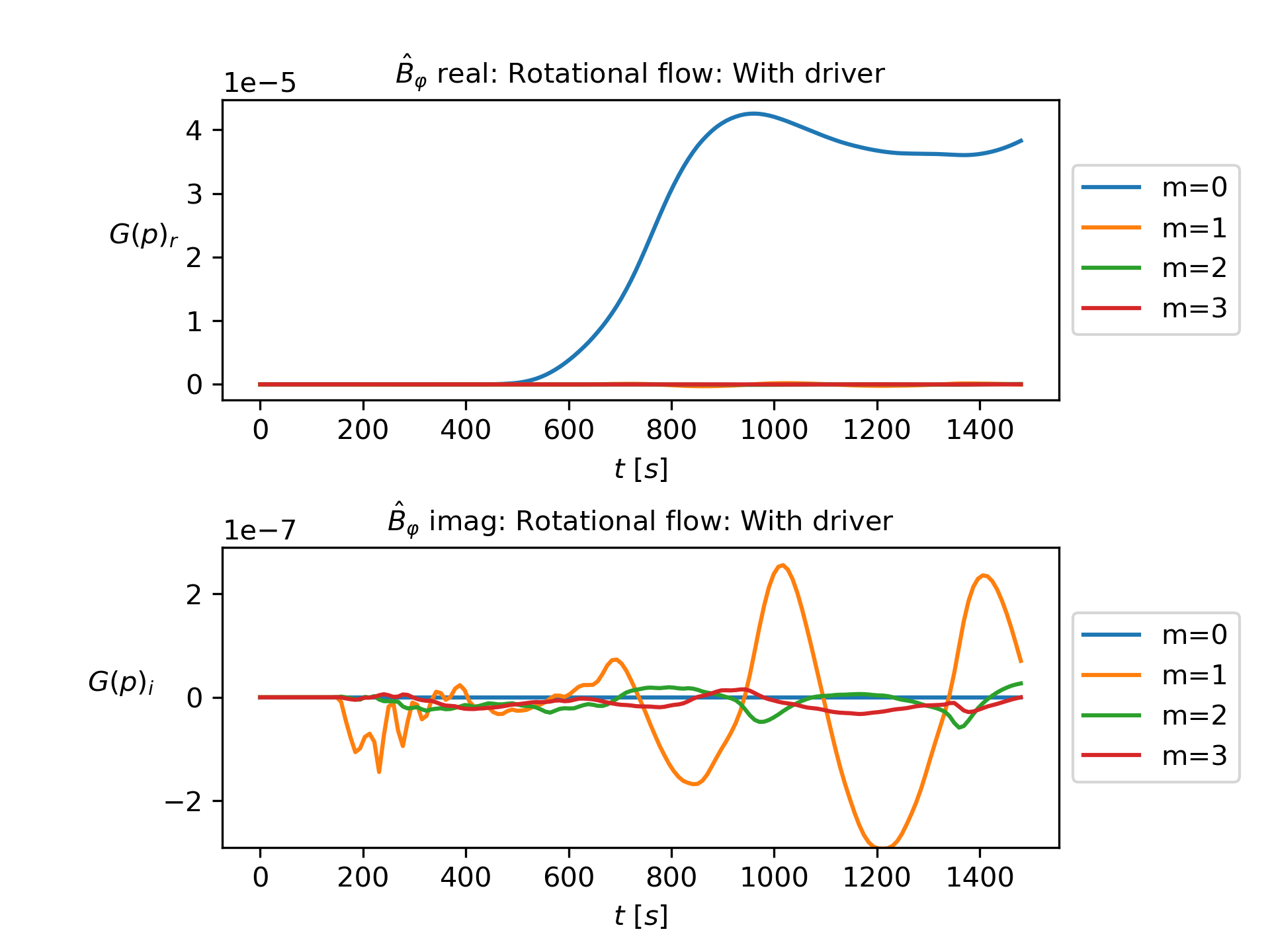}
    \caption{}
    \label{fig:rotflow_bphi_FFT}
\end{subfigure}
\caption{Real and imaginary Fourier coefficients of $\hat{B}_r$ and $\hat{B}_{\varphi}$ in the corona ($z=-23.14$ Mm) around flux surfaces at a radial location of $r=7.20$~Mm. Panels (a) and (b) show the Fourier coefficients for the simulation of a rotational flow applied but with no photospheric wave driver. However, panels (c) and (d) display the magnetic field Fourier coefficients for the simulation with a rotational flow and an additional photospheric wave driver.}
\label{fig:fourier_coefficients_magfield}
\end{figure*}

Further evidence of this mode being interpreted as a torsional Alfv\'{e}n mode arising from the non-steady flow applied at the lower boundary can be determined by examining the Fourier coefficients of the magnetic field perturbation. If a torsional Alfv\'{e}n wave/pulse were excited, then we should also expect to see a similar signal in the axisymmetric mode from the magnetic field perturbation. Figure \ref{fig:fourier_coefficients_magfield} displays the Fourier coefficients of the $\hat{B}_r$ and $\hat{B}_{\varphi}$ components for the simulations in the presence of a rotational flow both with and without an additional photopsheric wave driver. Figure \ref{fig:fourier_coefficients_magfield} shows the different periodicities which arise from the applied rotational flow and the implemented wave driver. The axisymmetric perturbations from the torsional Alfv\'{e}n pulses generated from the non-steady rotational flow are evident in the decomposition of the azimuthal magnetic field perturbation. The signals in both the $\hat{v}_{\varphi}$ and $\hat{B}_{\varphi}$ (real) coefficients for the axisymmetric mode appear at the same moment in time ($t=500$ s) for the simulation with a rotational flow and a wave driver, suggesting that they arise from the same formation mechanism.

\subsection{Transport of Poynting flux}\label{subsec:poyntingflux}

\begin{figure}
\centering
\includegraphics[width=0.45\textwidth]{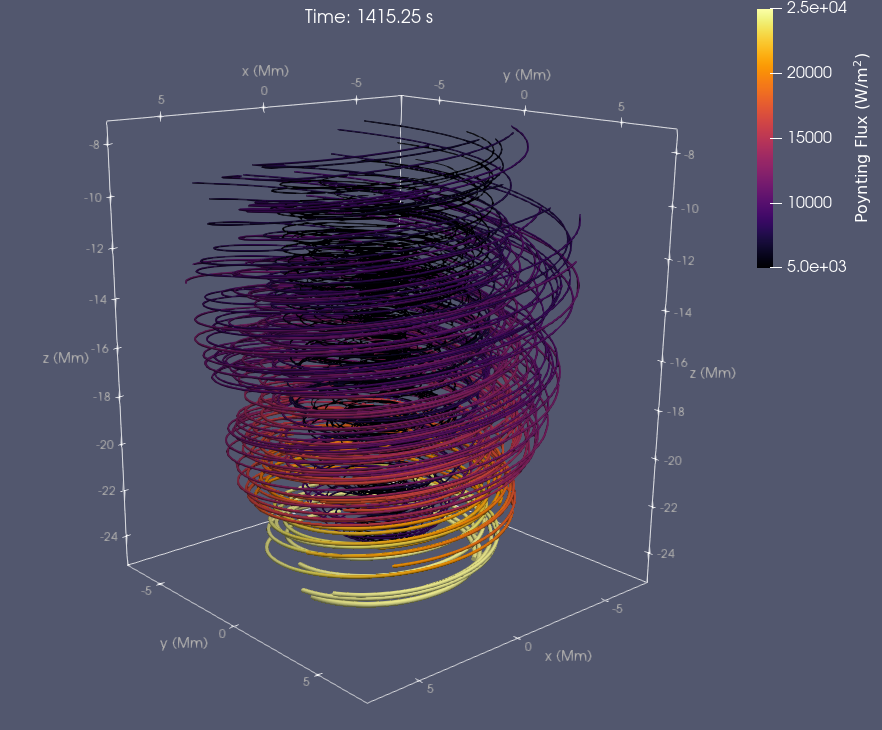}
\caption{Snapshot at $t=1415$~s of the 3D streamlines of the total Poynting flux in a box of the simulation domain. The magnitude of the Poynting flux is visualised through both colour and width of the 3D tube.}
\label{fig:3D_Poynting}
\end{figure}

\begin{equation}\label{eqn:PoyntingFlux}
    \mathbf{S} = -\frac{1}{\mu_0}(\mathbf{v}\times\mathbf{B})\times\mathbf{B}.
\end{equation}

As the Poynting flux, expressed in Equation (\ref{eqn:PoyntingFlux}) with $\mu_0$ the magnetic permeability of free space, provides an indication on the magnetic energy transport, it would be extremely beneficial to understand the manner in which the Poynting flux is transported through the simulated vortex tube and, if possible, to provide some insight into the transport of Poynting flux from MHD waves guided by the structure. Figure \ref{fig:3D_Poynting} provides a snapshot near the end of the simulation of streamlines of the total Poynting flux in the vortex tube. It is evident that there is a large amount of Poynting flux near the base of the vortex tube which is understandable as this is the location of the rotational flow at the bottom boundary. Moreover, the streamlines of the Poynting flux follow a helical trajectory in the corona, whereas the streamlines are more circular and horizontal below the transition region. A significant amount of Poynting flux is still present in the corona ($\approx 10$ kW/m$^2$), however, this figure displays the total Poynting flux in the simulation containing contributions from the background motions and the Poynting flux carried by perturbations. It should be emphasised that a large majority of the total Poynting flux in the simulation is a result of the background flow incorporated in the simulation (i.e. can be attributed to the rotating plasma itself).

\begin{figure*}
\centering
\includegraphics[width=0.95\textwidth]{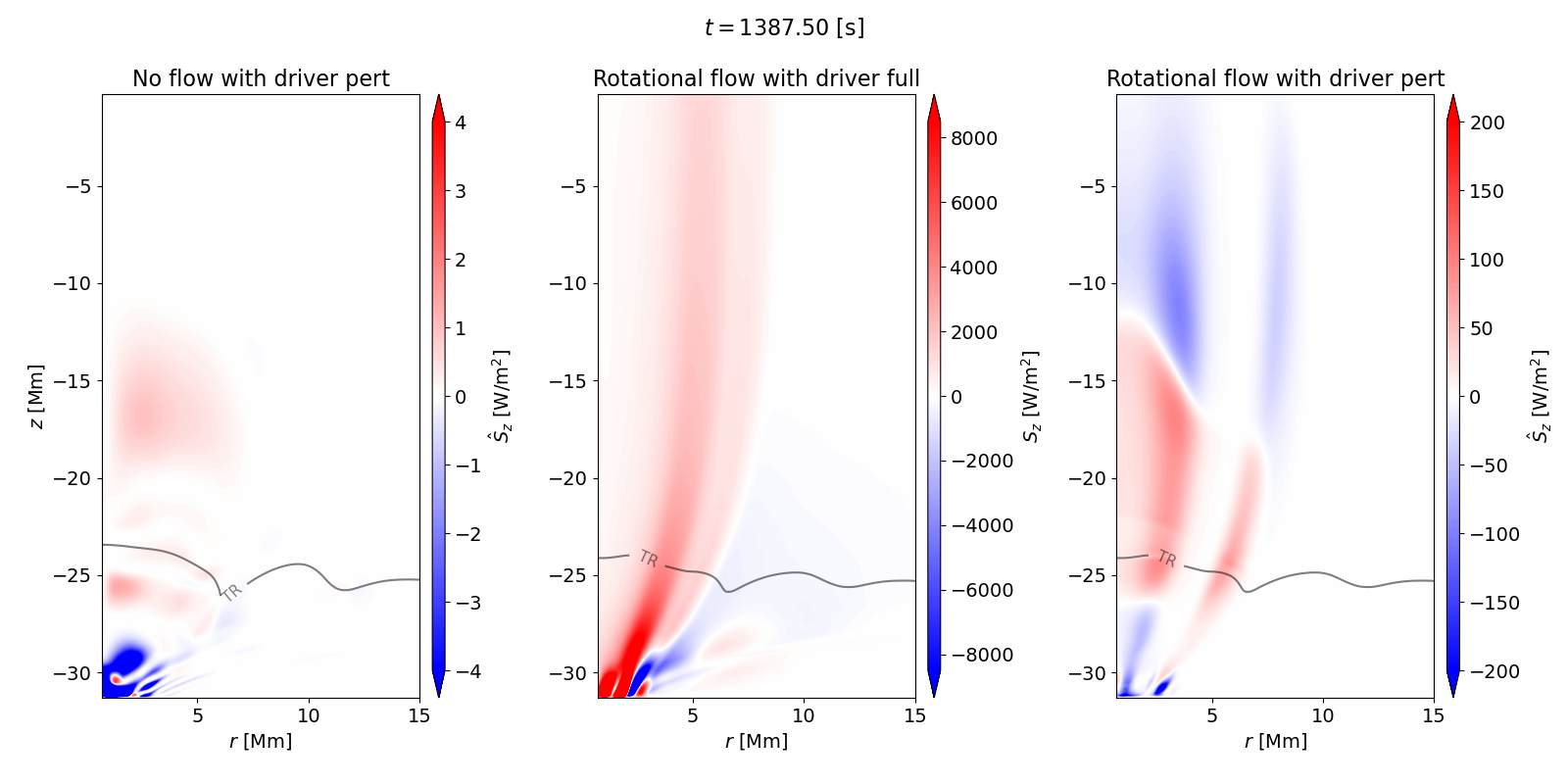}
\caption{Vertical cuts for a snapshot at $t=1388$~s at an angle of $\varphi=\pi/2$ for the perturbed (`pert' in subtitle) vertical component of Poynting flux $\hat{S}_z$ in the simulation with no rotational flow (left panel). Middle panel: The total (`full' in subtitle) vertical component of the Poynting flux, $S_z$ for the simulation with a rotational flow and a wave driver. Right panel: the perturbed vertical component of Poynting flux, $\hat{S}_z$, for the simulation with a rotational flow and a wave driver. The height of the transition region (TR) is highlighted by the grey contour and the colourbar is saturated in all plots. }
\label{fig:Poynting_comparison}
\end{figure*}

Figure \ref{fig:Poynting_comparison} displays vertical cuts of the $z$ component of the Poynting flux in the simulations with a photospheric driver both with and without a background rotational flow present. The left hand panel of Figure \ref{fig:Poynting_comparison} shows the perturbed vertical component of the Poynting flux, $\hat{S}_z$, resulting from the waves driven by the driver applied locally at the lower boundary. It is evident that there is some leakage of magnetic energy below the transition region in the chromosphere and that very little Poynting flux is channelled into the corona. This was a main conclusion by \citet{Skirvin2024modeconv} (see also \citet{Riedl2021}) who demonstrated that the majority of the energy flux is contained within the hydrodynamic component due to the nature of the driver and that wave energy is lost by lateral leakage in the chromosphere. 

The middle panel of Figure \ref{fig:Poynting_comparison} displays the total vertical component of the Poynting flux, $S_z$, in the simulation with a rotational flow and a wave driver. It is immediately apparent that there is a significantly greater amount of unsigned Poynting flux produced due to the rotational perturbation at the photosphere, and that this energy is channelled into the corona through the vortex tube. The right hand panel of Figure \ref{fig:Poynting_comparison} displays the perturbed vertical component of the Poynting flux, $\hat{S}_z$, for the simulation with a rotational flow and a photospheric wave driver. A clear propagating signal is observed and, strikingly, a remarkable increase in Poynting flux carried by perturbations is channelled into the corona. Not only is the magnitude of vertical Poynting flux associated with wave propagation greater in the presence of the vortex tube, the magnetic energy is channelled a greater distance into the coronal volume, highlighting the efficiency of solar vortex tubes to channel wave energy flux. This is attributed to the dynamical pressure resulting from the vortex flow, creating an overdense boundary layer which traps waves inside the vortex tube. For comparison, the maximum absolute value of $S_z$ in the left hand panel of Figure \ref{fig:Poynting_comparison} is $176$~W m$^{-2}$, whereas the maximum absolute value of $S_z$ in the right hand panel of 
Figure \ref{fig:Poynting_comparison} is $386$~W m$^{-2}$. Therefore, at the foot-point, the ratio between the maximum value of $S_z$ in the two simulations (with and without the rotational flow) is roughly a factor of $2$. However, it is clear that this ratio is different in the corona, where there is much greater $S_z$ in the simulation with a rotational flow, highlighting the effectiveness of the vortex tube at transporting magnetic energy flux. We can estimate the transmission of $S_z$ into the corona, relative to the photosphere, for each simulation when the wave driver is present. From Figure \ref{fig:Poynting_comparison}, we can conclude that the unsigned $S_z$ in the corona when no rotational flow is included is of the order $~1$~W m$^{-2}$. On the other hand, when a rotational flow is present, the unsigned $S_z$ in the corona is roughly $50-100$~W m$^{-2}$. This results in a transmission coefficient of $0.6\%$ for the simulation with no rotational flow included, and a transmission coefficient of $13-26\%$ when a rotational flow is present.

\begin{figure*}
\centering
\includegraphics[width=0.95\textwidth]{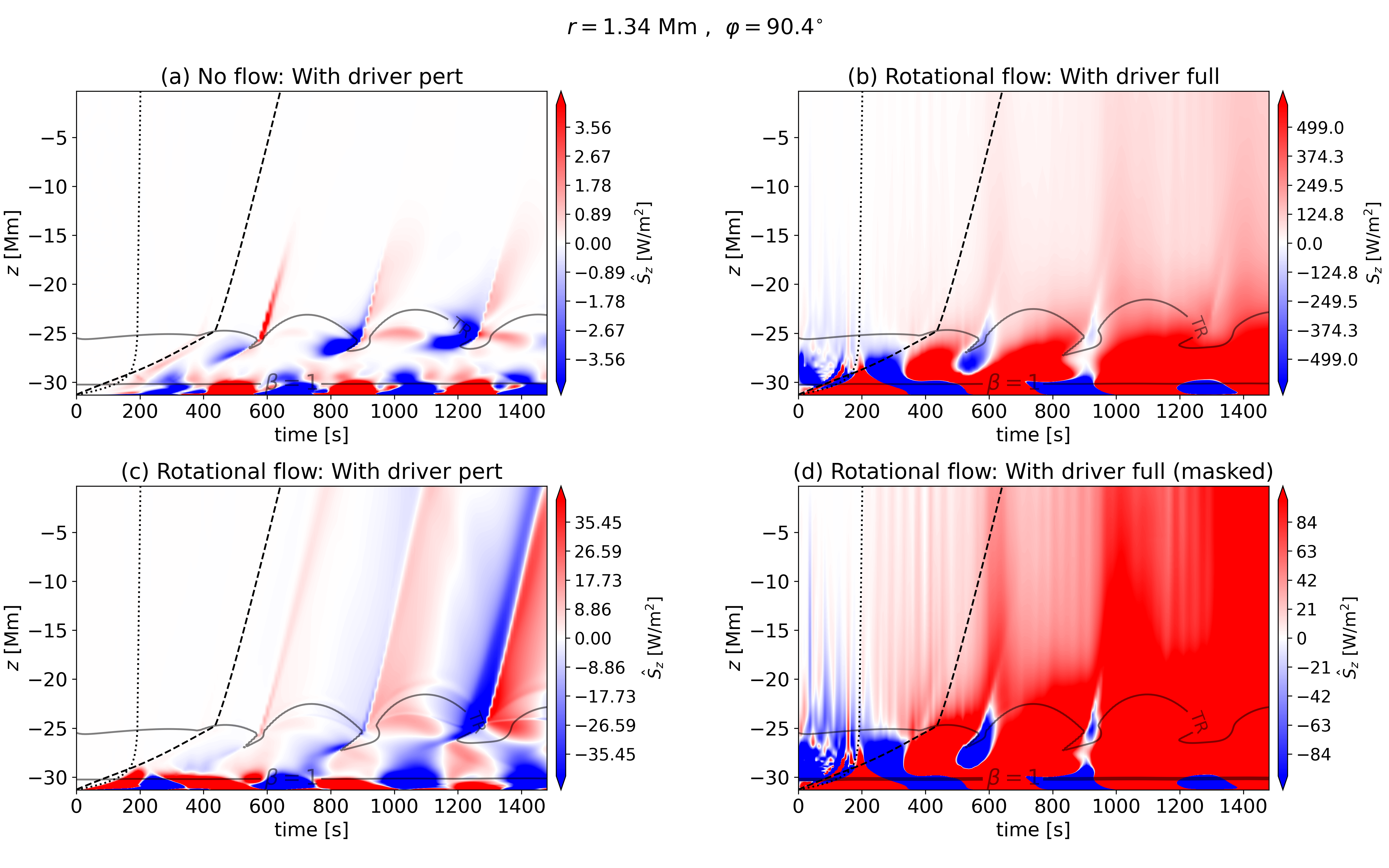}
\caption{Time distance plots at a radial location of $r=1.34$~Mm and an azimuthal angle of $\varphi=90.4^{\circ}$ of the vertical Poynting flux. Panel (a) shows the perturbed vertical Poynting flux, $\hat{S}_z$, in the simulation without a background rotational flow but with a wave driver \citep[see e.g.][]{Skirvin2023ApJ, Skirvin2024PDs}. Panel (b) displays the full vertical Poynting flux $S_z$ in the simulation with a background rotational flow and a wave driver. Panel (c) highlights the perturbed vertical Poynting flux in the simulation with a rotational flow and a wave driver (i.e. after the rotational flow simulation is subtracted). Panel (d) is the same as panel (b) but values with absolute magnitude greater than $100$ W m$^{-2}$ have been masked. The dotted line shows the trajectory of the local Alfv\'{e}n speed, whereas the dashed line highlights the sound speed. The height of the transition region (TR) and the plasma-$\beta=1$ layer are denoted by the labelled contours and the colourbar is saturated in all plots.}
\label{fig:Poynting_comparison_tds}
\end{figure*}

It could be assumed that the strong increase in perturbed vertical Poynting flux may be associated with the torsional Alfv\'{e}n waves excited by the non-stationary rotational motion at the foot-point. However, Figure \ref{fig:Poynting_comparison_tds} does not suggest that the Poynting flux propagates at the Alfv\'{e}n speed, instead, the vertical Poynting flux propagates at a speed closer to the local sound speed suggesting that the $\hat{S}_z$ is carried by slow magnetoacoustic waves. On the other hand, there is a coupling between slow magnetoacoutsic and Alfv\'{e}n waves relating to physical effects such as field line curvature, resonances and shocks \citep{hollweg1971, Ivanov1976,crouch2005} which may cause the Alfv\'{e}n waves to have an indirect effect on the generated $\hat{S}_z$ propagating with the local sound speed.

Recent work by \citet{Skirvin2024Poynting} have shown that MHD waves carry increased amounts of $\hat{S}_z$ when in the presence of a background rotational flow. They demonstrate how the increase in Poynting flux associated with the waves depends on the amplitude ratio between the strength of the background flow and the strength of the perturbation. From Figure \ref{fig:vphi_with_time}, we can assume that the amplitude ratio $v_0/\hat{v}$ = 5-10, where $v_0$ denotes the background flow strength and $\hat{v}$ denotes the strength of the perturbation. This amplitude ratio should correspond to an increase of roughly $\hat{S}_z$ of $10-20\%$ for sausage modes and $20-100\%$ for kink modes. It is evident that the perturbed Poynting flux present in the right hand panel of Figure \ref{fig:Poynting_comparison} is much greater than this predicted increase, suggesting the presence of other MHD modes in the simulation. On the other hand, the increase in $S_z$ may also be due to the waves propagating in a different plasma environment when the rotational flow is included, as highlighted in Figure \ref{fig:transverse_profiles}. It is possible that the Poynting flux associated with sausage and kink modes is increased in the simulation when compared to a scenario without a rotational flow present, however, accurately tracing the Poynting flux transported by these exact modes is extremely complex to isolate.

Figure \ref{fig:Poynting_comparison_tds} displays time-distance plots of the vertical component of the Poynting flux associated with the perturbations (waves), $\hat{S}_z$, in the simulations both with and without the inclusion of a background rotational plasma flow. In these plots, the trajectories of the local Alfv\'{e}n and sound speeds are shown and it is clear that in both simulations, the $\hat{S}_z$ propagates at the local sound speed in a low-$\beta$ plasma. The time-distance plots also demonstrate how the magnitude of $\hat{S}_z$ increases when in the presence of a background rotational flow. Panel (b) in Figure \ref{fig:Poynting_comparison_tds} displays the full value for the vertical Poynting flux in the simulation with a rotational flow and a wave driver, the same quantity is shown in panel (d) where values with absolute magnitude greater than $100$~W/m$^2$ have been masked. In this plot, the initial background motions are evident as Poynting flux is present before waves have entered the coronal volume, however, it is also clear that waves are generated by the rotational flow appearing at later times and propagate with the local Alfv\'{e}n velocity. There is a high frequency component present in the $S_z$ signal which propagates at the Alfv\'{e}n speed and appears to be an Alfv\'{e}n wave/pulse whose origin is unclear. This signal is observed at smaller radii where the magnetic field at the foot-point is stronger and the transverse inhomogeneity is larger. As this high frequency signal is not present in panel (c) after the rotational flow has been subtracted, it suggests that the high-frequency $S_z$ signal travelling at the Alfv\'{e}n speed is a non-linear response of the interaction between the rotational flow and the stratified flux tube. 

Panel (c) in Figure \ref{fig:Poynting_comparison_tds} shows the perturbed vertical component of the Poynting flux (i.e. after the rotational flow has been subtracted). Interestingly, $\hat{S}_z$ shows extremely similar characteristics to propagating disturbances (PDs) associated with the transition region dynamics \citep{Samanta2015, Skirvin2024PDs}. This has recently been discussed by \citet{Skirvin2024PDs} who demonstrated that the PDs are predominantly acoustic in nature, propagate along the magnetic field at the local sound speed, and, as a result can be associated with slow magnetoacoustic waves. They have been shown to carry into the corona sufficient hydrodynamic energy flux ($>100$~Wm$^{-2}$) to contribute to the heating of local plasma, however, in the presence of a background plasma flow the PDs also carry the same order of magnetic energy flux into the corona which is injected directly at the transition region. Whilst slow magnetoacoustic waves carry zero net Poynting flux in the linear regime, it is evident that the unsigned $\hat{S}_z$ associated with these slow magnetacoustic waves is larger when a background (rotational) plasma flow is present.

Both panel (b) and (d) in Figure \ref{fig:Poynting_comparison_tds} highlight the positive vertical Poynting flux associated with the background rotational flow and the waves, generated through the interaction of the loop with the flow itself, are transporting magnetic energy upwards into the corona. The positive sign of the Poynting flux can be understood by looking at Equation (33) from \citet{Skirvin2024Poynting} and repeated below:
\begin{equation}\label{eqn_zero_oredr_Sz}
[(\mathbf{v_0}\times\mathbf{B_0})\times\mathbf{B_0}]_z = (\underbrace{B_{0,\varphi} v_{0,z}}_{\text{T1}} - \underbrace{v_{0,\varphi}B_{0,z}}_{\text{T2}})B_{0,\varphi}.
\end{equation}
Equation (\ref{eqn_zero_oredr_Sz}) represents the vertical component of the Poynting flux associated with background motions only (i.e. zeroth order), subscripted with $0$. There are two terms inside the bracket which, when combined with the sign of the background magnetic twist $B_{0,\varphi}$, determine whether the Poynting flux is directed upwards or downwards.
\begin{figure}
\centering
\includegraphics[width=0.45\textwidth]{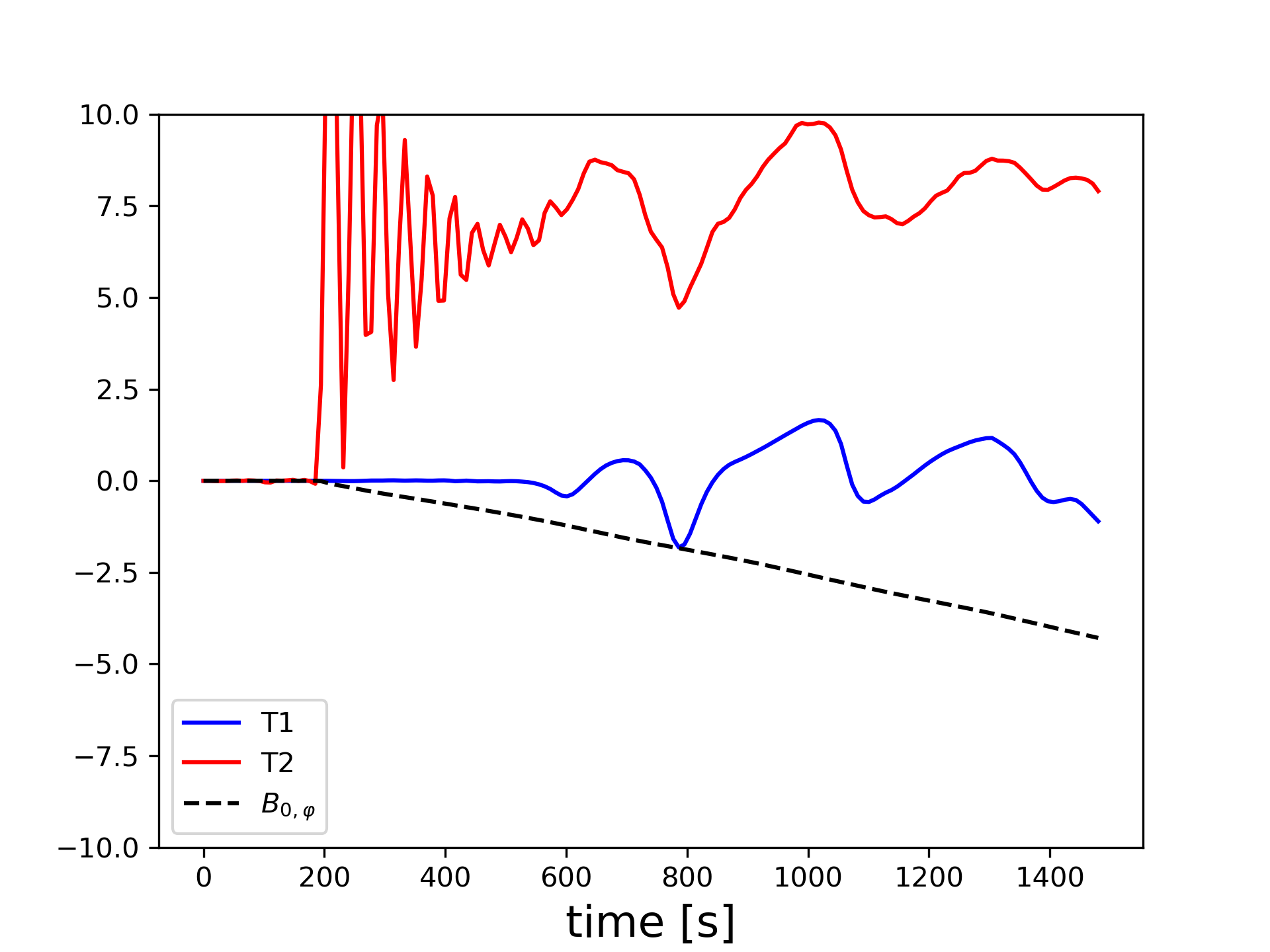}
\caption{The numerical values of the two terms, T1 (blue) and T2 (red) inside the bracket of Equation (\ref{eqn_zero_oredr_Sz}) over time. Also shown is the background magnetic twist $B_{0,\varphi}$ in units of Gauss. This figure is produced at a fixed position of $r=1.34$~Mm, $\varphi=\pi/2$ and $z=-23.14$~Mm.}
\label{fig:Poynting_terms_comparison}
\end{figure}
The numerical values of these two terms in Equation (\ref{eqn_zero_oredr_Sz}) are displayed in Figure \ref{fig:Poynting_terms_comparison}. It is evident that term 2 is greater than term 1 throughout the duration of the simulation, therefore, the overall sign of the terms inside the bracket is negative. However, the value of the background magnetic twist is negative, which results in a positive vertical Poynting flux, associated with the rotational flow. If the rotational flow is introduced in the opposite direction, then $B_{0,\varphi}$ will be positive, but term 2 in Equation (\ref{eqn_zero_oredr_Sz}) will become negative, as $v_{0,\varphi}$ will change sign. Therefore, this scenario will still result in a positive $S_z$, regardless of the direction of the rotational flow at the photosphere, because term 2 is always greater than term 1. This is only valid when term 1 plays a small role, in the case when $B_{0,\varphi} \ll B_{0,z}$, as is true for the current simulation. The implications in a strongly twisted tube is a different story and it may no longer be possible to produce a net upwards Poynting flux in strongly twisted tubes.

\section{Discussion \& Conclusions}\label{sec:conclusions}

In this study we have performed a 3D numerical simulation aimed to generate a solar vortex tube and drive various MHD waves at the photosphere. A combination of sausage, kink and torsional Alfv\'{e}n waves are supported by the vortex tube as a result of both the photospheric wave driver and the non-stationary nature of the photospheric vortex flow. These waves are present in the simulation arising from different physical mechanisms - all of which are feasible in the solar atmosphere. The tube modes (e.g. kink and sausage) are produced by perturbations at the photosphere from p-modes which are absorbed by the magnetic flux tube, whereas the torsional Alfv\'{e}n modes are generated by the rotational flow at the foot-point which is non-stationary (i.e. has a non-zero time derivative).

We have demonstrated that it becomes difficult to accurately decompose different MHD modes when background rotational flows are present, and that some wave signatures may be misleading when compared to classical understanding of waves in a uniform magnetic flux tube. For example, the sausage mode and torsional Alfv\'{e}n mode display very similar characteristics when their motions are decomposed relative to circular contours of constant magnetic field strength \citep{Giag_2015}. Furthermore, we have presented a Fourier analysis of the azimuthal wavenumbers to highlight the disparity between the different Fourier coefficients, unlike the case when no background rotational flow is present. The FFT analysis underlined the various MHD modes present in the simulation, however, provided a clearer insight into the contribution of sausage, kink and torsional Alfv\'{e}n modes to the various azimuthal signals of different velocity components. The numerical model presented in the current study, along with the incorporated rotational flow, provides a highly non-uniform and structured plasma, where identifying pure wave modes (e.g. slow, fast, Alfv\'{e}n) using velocity field perturbation information becomes increasingly complex as the modes become coupled in such systems.

We have examined the Poynting flux carried by waves present in both simulations (with and without a background rotational flow). Our results indicate that the Poynting flux transported by MHD waves is greatly increased when propagating within a vortex tube compared to the simulation where no rotational flow is considered. As discussed in Section \ref{subsec:vortex_formation}, the presence of a vortex flow increases the dynamical pressure at the boundary of the vortex tube, which creates a layer of dense plasma which acts as an effective waveguide to trap and guide wave energy. Furthermore, in the presence of rotational flow, we have shown that the Poynting flux has a helical structure 
\citep[as reported by][]{Silva2024}, highlighting the need for careful analysis when trying to interpret the Poynting flux present in observations where all three components of the velocity and magnetic field may not be retrievable. There is clear upward Poynting flux transported by the Alfv\'{e}n waves generated from the rotational flow at the photosphere. The perturbed vertical Poynting flux associated with the wave driver travels at the local sound speed in the corona and shares striking resemblance to propagating disturbances discussed by \citet{Skirvin2024PDs}, which may be connected to the Poynting flux connected with jets and compressible waves in the solar wind \citep{Bale2019}. Whilst the magnitude of the unsigned flux carried by slow magnetoacoustic waves is much larger when a background rotational flow is present, future work should investigate the net flux carried by these waves which would be available for heating in the upper atmosphere. Future work should investigate the energy transported by individual modes (e.g. sausage/kink). This analysis will be complex as the presence of a rotational flow will couple different azimuthal modes. Alternative diagnostics should be employed to study the modes, and their energies, present in complex plasma configurations \citep[see e.g.][]{Raboonik2024b, Raboonik2024}.

In the future, it is essential to understand the Poynting vector associated with MHD waves in the presence of background plasma flows, especially in the corona where direct measurement is extremely difficult and the solar wind is non-negligible. This will allow for accurate diagnosis of the wave energy flux available in, for example, transverse kink waves, which is commonly computed using a 1D approximation \citep{Tomczyk_et_al_2007, McIntosh_et_al_2011}. More detailed numerical simulations of MHD waves supported by vortex tubes, which are ubiquitous, in a realistic solar atmosphere are required to correctly diagnose the energy flux transported along the magnetic field in a 3D geometry. Further analytical and numerical investigations are needed to understand the effect of background flows on the magnetic energy flux carried by different MHD modes. This is essential in the solar corona where these waves are in the presence of the solar wind which may affect the available energy carried by these waves.

\begin{acknowledgments}
We are grateful to the referee for their thoughtful suggestions which have greatly improved the quality of this manuscript. SJS and VF are grateful to the Science and Technology Facilities Council (STFC) grants ST/V000977/1, ST/Y001532/1.  VF, GV and IB thank The Royal Society, International Exchanges Scheme, collaboration with Instituto de Astrofisica de Canarias, Spain (IES/R2/212183), International Exchanges Scheme (NSTC) collaboration with National Central University, Taiwan (IEC/R3/233017), Institute for Astronomy, Astrophysics, Space Applications and Remote Sensing, National Observatory of Athens, Greece (IES/R1/221095), and Indian Institute of Astrophysics, India (IES/R1/211123) for the support provided. VF would like to thank the International Space Science Institute (ISSI) in Bern, Switzerland, for the hospitality provided to the members of the teams on `The Nature and Physics of Vortex Flows in Solar Plasmas' and `Tracking Plasma Flows in the Sun’s Photosphere and Chromosphere: A Review \& Community Guide'. VF and GV are grateful to the Institute for Space-Earth Environmental Research (ISEE, International Joint Research Program, Nagoya University, Japan) for the support provided.
\end{acknowledgments}

%


\bibliography{ref}{}
\bibliographystyle{aasjournal}

\end{document}